\documentclass[]{aa}
\usepackage[varg]{txfonts}

\usepackage{epsf,rotating,latexsym,amssymb} 
\usepackage{epsfig}   
\usepackage{epstopdf} 
\usepackage{rotating,latexsym,amssymb}           
\usepackage{mathptmx} 
\usepackage{graphicx}
\usepackage{times}
\usepackage{natbib}
\usepackage{lineno}       
\usepackage{color}

\newcommand{\kms}{$\mathrm { km\,s^{-1}}$}

\newcommand{\honda}{45P/H-M-P}
\newcommand{\denning}{72P/D-F}
\newcommand{\finlay}{15P/Finlay}
\newcommand{\machholz}{141P/Machholz 2-A}
\newcommand{\hondaa}{45P/H-M-P }
\newcommand{\denningg}{72P/D-F }
\newcommand{\finlayy}{15P/Finlay }
\newcommand{\machholzz}{141P/Machholz 2-A }

\begin{document}
\bibliographystyle{aa}

\title{Helios spacecraft data revisited: Detection of cometary meteoroid trails by in-situ dust impacts} 

\titlerunning{Helios Detections of Cometary Dust Trails in the Inner Solar System}
\authorrunning{H. Kr\"uger et al.}

\author{
Harald Kr\"uger\inst{1,2}  \and 
Peter Strub\inst{1,3} \and
Max Sommer\inst{3} \and
Nicolas Altobelli\inst{4} \and
Hiroshi Kimura\inst{2} \and 
Ann-Kathrin Lohse\inst{1} \and \\
Eberhard Gr\"un\inst{5,6} \and 
Ralf Srama\inst{3,7} 
}

\offprints{Harald Kr\"uger, krueger@mps.mpg.de}

\institute{
Max-Planck-Institut f\"ur Sonnensystemforschung, G\"ottingen, Germany  \and
Planetary Exploration Research Center, Chiba Institute of Technology, Narashino, Japan \and
Institut f\"ur Raumfahrtsysteme, Universit\"at Stuttgart, Germany \and
European Space Agency, European Space Astronomy Center, Madrid, Spain \and
Max-Planck-Institut f\"ur Kernphysik, Heidelberg, Germany \and
LASP, University of Colorado, Boulder, CO, USA \and
Baylor University, Waco, TX, USA
}

\date{Date received, date accepted}

\abstract
{Cometary meteoroid trails exist in the vicinity of comets, forming fine structure of the interplanetary dust cloud.
The trails consist predominantly of the largest cometary particles (with sizes of approximately 0.1 mm to 1 cm) which are 
ejected at low speeds and 
remain very close to the comet orbit for several revolutions around the Sun. In the 1970s 
two Helios spacecraft were launched towards the inner solar system. The spacecraft were equipped with in-situ
dust sensors which measured the distribution of interplanetary dust in the inner solar system for the first time. 
When re-analysing the Helios data, \citet{altobelli2006}
recognized a clustering of seven impacts, detected by Helios in a very narrow region of space at a true
anomaly angle of  $135\pm 1^{\circ}$, which the authors considered as
potential cometary trail particles. At the time,
however, this hypothesis could not be studied further.
}
{We re-analyse these candidate cometary trail particles in the Helios dust data to investigate the 
possibility that some or all of them indeed  originate from 
cometary trails and we constrain their source comets.}
{
The Interplanetary Meteoroid Environment for eXploration (IMEX) dust streams in 
space model is a new universal model for 
cometary meteoroid streams in the inner solar system, 
developed by \citet[][]{soja2015a}. We use IMEX to study cometary trail traverses by  Helios. 
}
{
During ten revolutions around the Sun, the Helios spacecraft intersected 13 
cometary trails. For the majority of these traverses the predicted dust fluxes are very low. In the narrow 
region of space where Helios detected the candidate dust particles, the spacecraft repeatedly traversed  
the trails of comets 45P/Honda-Mrkos-Pajdu\v{s}{\'a}kov{\'a} and 72P/Denning-Fujikawa with relatively high predicted 
dust fluxes. 
The analysis of the detection times and particle impact directions shows that four detected particles are 
compatible with an origin from these two comets. By combining measurements 
 and simulations we find a dust spatial density   
in these trails of approximately $\mathrm{10^{-8}\,m^{-3}}$ to $\mathrm{10^{-7}\,m^{-3}}$.
}
{The identification of potential cometary trail particles in the Helios data greatly benefitted from the 
clustering of trail traverses in a rather narrow region of space. The in-situ detection and  analysis of 
meteoroid trail particles which can be traced back to their source bodies by spacecraft-based dust analysers opens a new 
window to remote compositional analysis of comets and asteroids without the necessity to fly a spacecraft  to or 
even land on those celestial bodies. This provides new science opportunities for future  missions
like Destiny$^+$, Europa Clipper and IMAP.
}

\maketitle
\section{Introduction}

A cometary dust tail consists of small sub-micrometer sized dust particles that are blown out by solar radiation 
pressure forces. Larger dust particles form the dust coma and later spread in the orbit of the comet as a 
result of small differences in orbital period. They form a tubular structure around the parent comet's orbit 
called a dust trail.
Dust trails in the vicinity of comets were first observed by the Infrared Astronomical 
Satellite \citep[IRAS;][]{sykes1986}. IRAS  identified  a 
total of eight  cometary meteoroid trails \citep{sykes1992}. In 
subsequent infrared observations  
at least 80\% of the observed Jupiter-family comets were associated with dust trails  which can thus be 
considered one of their generic features \citep{reach2007}. More
recently, detections of dust trails were also reported in the visible wavelength range \citep{ishiguro2007}.
A recent review about cometary dust including dust trails was  given by \citet{levasseur-regourd2018}.

These trails form fine-structure superimposed upon the interplanetary 
background dust cloud. They consist of the largest cometary particles \citep[with sizes of approximately 
0.1 mm to 1~cm;][]{agarwal2010}, which are ejected at low speeds and remain very close to the comet 
orbit for several revolutions around the Sun. Trail particles are much bigger than the 
particles in the comet's dust tail, and the latter disperse more rapidly as a result of higher 
ejection speeds and solar radiation pressure. 
When the Earth intercepts a cometary trail, the
 particles collide with the atmosphere and show up as meteors and fireballs \citep[][and references therein]{koschny2019}. 
 Effects of meteoroid impacts were also observed on the Earth Moon and on other planets \citep{christou2019}.
Up to now there is no known detection of a cometary trail with a spacecraft-based in-situ dust detector. 

In the 1970s two Helios spacecraft were launched towards the inner solar 
system. The goal of the missions was to reach an orbital 
perihelion at 0.3~AU from the Sun  (Figure~\ref{fig:traj_1}), performing measurements of the 
interplanetary magnetic field, 
the solar wind, cosmic radiation, the zodiacal light, and the interplanetary dust distribution.
The spacecraft were equipped with two in-situ dust sensors each, which  measured the distribution of
interplanetary dust in the inner solar system for the first time \citep{gruen1980a,gruen1981b}. 

\citet{altobelli2006} re-analysed the Helios dust data  searching for interstellar impactors \citep{gruen1994a,krueger2019b}. The
authors recognized a cluster of seven impacts in a very narrow range of the spacecraft's true anomaly angle.
Remarkably, these impacts were detected during a total of six Helios orbits at almost exactly the same spatial 
location. This coincidence led the authors to speculate that the impacts may have occurred during repeated 
spacecraft traverses of a cometary meteoroid trail. 
At the time, however, no detailed cometary
trail model for the inner solar system was available to further investigate this hypothesis.

Recently, the Interplanetary Meteoroid Environment for eXploration (IMEX) dust streams in 
space model was developed by \citet{soja2015a,soja2015b} under contract by the European Space Agency. 
It is a new universal model that simulates recently created cometary dust trails in the 
inner solar system. The IMEX model follows the trails 
of 420 comets. 
IMEX is a physical model for dust dynamics and orbital evolution. It
is ideal for studying meteor streams and cometary dust trails as measured by in-situ detectors 
and observed in infrared images.

In this work, we use the IMEX model to investigate cometary trail traverses by the Helios spacecraft. We
compare the  measurements of  seven candidate cometary trail particles identified by \citet{altobelli2006} 
with trail traverses predicted by the model in order 
to investigate the hypothesis that a few or all of these particles  originated from a cometary meteoroid trail.
In Section~\ref{sec:mission} we briefly describe the Helios mission and the dust instruments on board, and 
in Section~\ref{sec:model} we summarise the IMEX model. We present the results of 
our IMEX simulations and compare them with the Helios measurements in 
Section~\ref{sec:results}. In Section~\ref{sec:flux} we constrain the dust fluxes in the
trails of two comets identified in our analysis. Section~\ref{sec:discussion}
is a discussion and Section~\ref{sec:perspectives} is an outlook to future perspectives. In Section~\ref{sec:conclusions} 
we summarize our conclusions.

\section{Helios Dust Measurements}

\label{sec:mission}

\begin{figure}[tb]
	\vspace{-0.4cm}
	\hspace{-11mm}
		\includegraphics[width=0.67\textwidth]{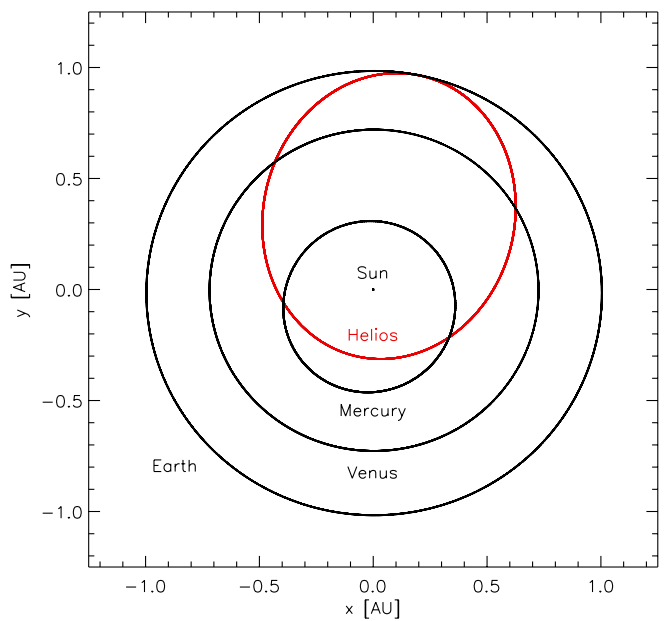}
		\vspace{-7.5cm}
	\caption{The orbits of Helios (red), Mercury, Venus and Earth. The X-Y plane is the 
	ecliptic plane with vernal equinox oriented towards the +X direction.
		  }
	\label{fig:traj_1}
\end{figure}

The Helios~1 spacecraft (we refer only to Helios~1 throughout this paper as the Helios~2 dust instruments
did not provide useful dust measurements because of high noise rates on board the spacecraft) 
 was launched into a 
heliocentric orbit on 10 December 1974. The Helios trajectory was in the ecliptic plane 
(inclination $\mathrm{i=0.02^{\circ}}$). 
The eccentricity 
of the elliptical orbit was about $e = 0.56$, the perihelion was located at 0.31~AU from the Sun,  the 
aphelion at 0.98~AU, and the argument of  perihelion was $258.4^{\circ}$. The spacecraft's orbital period 
was about 190~days. The Helios orbit is shown in Figure~\ref{fig:traj_1}.
 
\begin{figure}[tb]
	\vspace{-0.6cm}
	\hspace{15mm}
		\includegraphics[width=0.33\textwidth]{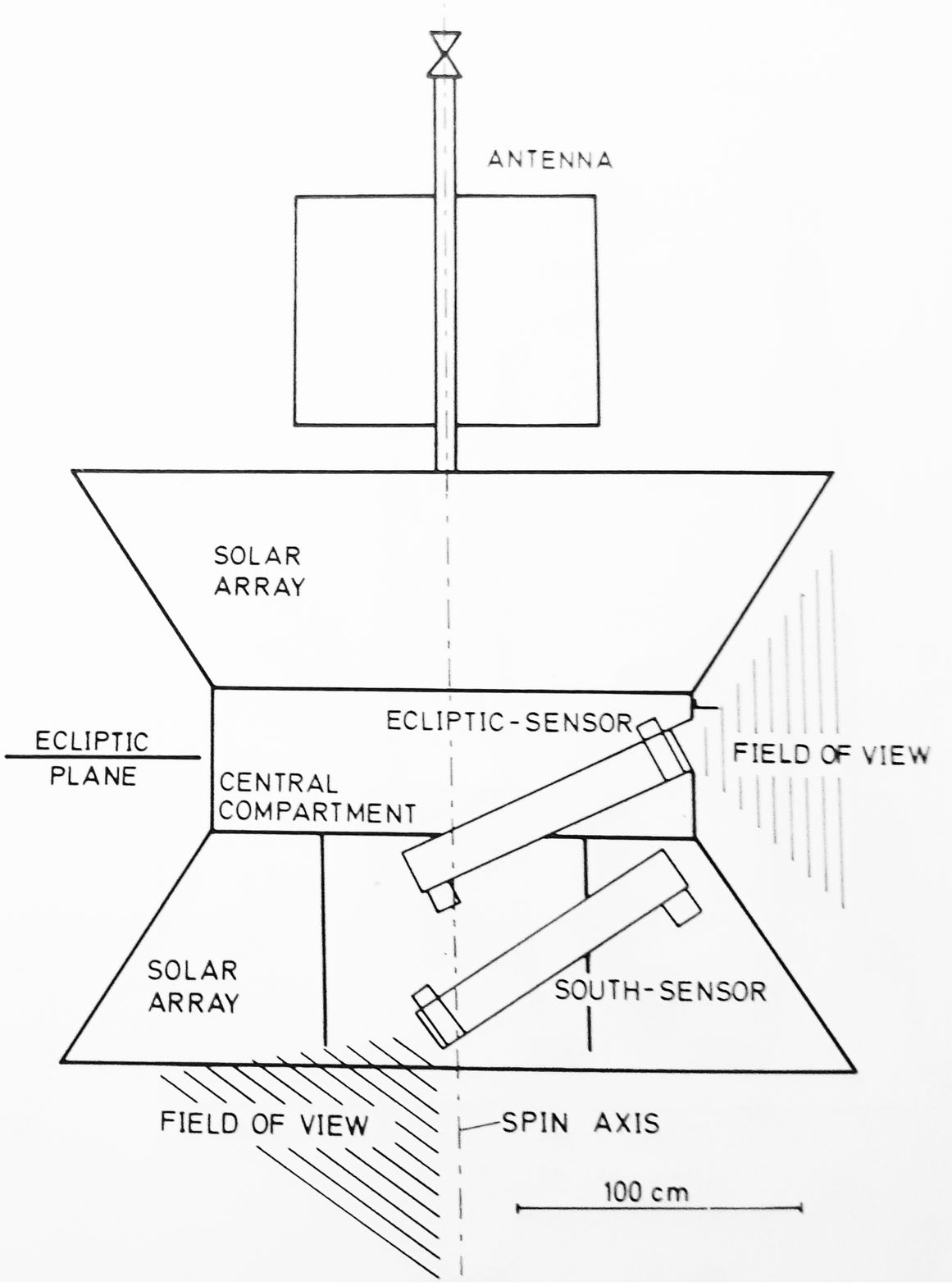}
		\vspace{-0.2cm}
	\caption{Schematic drawing of the Helios spacecraft, carrying the twin dust sensors. The ecliptic sensor 
	was sensitive to dust particles on low inclined orbits, while the south sensor measured 
	particles coming from the ecliptic south direction.
	Adapted from \citet{altobelli2006}.
		  }
	\label{fig:helios}
\end{figure}

The spacecraft was spin-stabilized with
a spin axis pointing normal to the ecliptic plane and a spin period of one second. 
In Figure~\ref{fig:helios} we show a schematic drawing of the spacecraft.
 It carried two dust instruments, the
ecliptic sensor which was exposed to sunlight, and the south sensor which was shielded  by the
spacecraft from direct sunlight \citep{dietzel1973,fechtig1978,gruen1980a,gruen1981b}. 
Between 19 December 1974 and 2 January 1980 the dust sensors transmitted the data of 235 dust impacts 
to Earth \citep{gruen1981b}. The true number of dust impacts onto the 
instruments was larger because of incomplete data transmission and instrumental dead-time \citep{gruen1980a}.

The measurement principle of the Helios dust instruments  was based on the impact ionization generated upon impact of a high-velocity 
projectile onto a solid target \citep{dietzel1973,auer2001}. From the measured signals, both the impact velocity and the 
mass of the impacting dust particle could be derived. 
In addition, the Helios instruments had time-of-flight mass spectrometer subsystems, providing 
information about the chemical elemental composition of the impactor. The target was a venetian blind consisting of gold strips 
held at ground potential. The constituents of the impact plasma were electrons, positive and negative ions, neutral atoms 
or molecules, and residual fragments of the impactor and target. Electrostatic fields separated the positive and 
negative charges generated during the impact. The particle impact speed and mass were derived from   both the rise time 
and the amplitude of the charge signals \citep{eichhorn1978a,eichhorn1978b,gruen1995a}. The Helios instruments 
had a detection threshold for dust particles with masses of approximately $\mathrm{3 \cdot 10^{-16}\,kg}$ 
at an impact speed of $\mathrm{10\,km\,s^{-1}}$ \citep{gruen1980a}.

 The two Helios dust sensors were twin instruments. The so-called south sensor was sensitive to 
dust particles on inclined prograde heliocentric orbits. For an observer on board the spacecraft, those particles came 
from the ecliptic south direction. The second sensor was called ecliptic sensor since its field-of-view pointed 
towards the ecliptic plane. The field-of-view of each sensor was a cone with
a half angle of $65^{\circ}$ (ecliptic sensor) and $73^{\circ}$ (south sensor), respectively, centered on the 
sensor axis  \citep{gruen1980a}. Both instruments were partially
shielded by the spacecraft structure, resulting in slightly different target areas: $\mathrm{54.5\,cm^2}$ for 
the ecliptic sensor, and $\mathrm{66.5\,cm^2}$ for the south sensor. 

As the ecliptic sensor pointed into 
the Sun once per spacecraft rotation, an additional aluminum-coated parylene foil of $\mathrm{0.3 \,\mu m}$ thickness 
covered the instrument aperture. This foil prevented solar radiation from entering the sensor and heating it 
up beyond safe operations but dust impactors could penetrate it. However, the sensitivity of the sensor was
decreased. In contrast, the south sensor had only a protection against the solar wind plasma, which did not 
decrease its sensitivity. 

The ecliptic sensor was sensitive to dust particles approaching with elevations from 
$-45^{\circ}$ to $+55^{\circ}$ with respect to the ecliptic plane. The south sensor could detect  particles 
with trajectory elevations from $-90^{\circ}$ (ecliptic south-pole) to $-4^{\circ}$. During one spin revolution 
of the spacecraft, both instruments scanned an entire circle along the ecliptic plane. More details about the
instruments and their calibration can be found in \citet{gruen1980a,gruen1981b} and \citet{altobelli2006}.

\begin{figure}[tb]
\vspace{-3.cm}
	\hspace{-12mm}
		\includegraphics[width=0.61\textwidth]{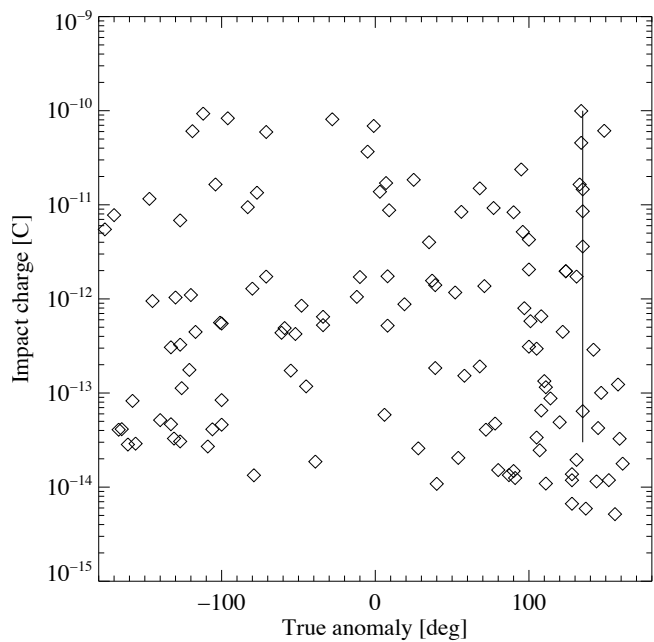}
	\vspace{-3.cm}
	\caption{Impact charges detected by the ecliptic and the south sensor as a function of Helios' true anomaly angle $\eta$
	for a subset of the Helios dust data, from \citet{altobelli2006}. 
	The vertical line shows a cluster of seven impacts which are candidates for  
	cometary trail particles detected when the spacecraft intercepted one or more cometary trails.
	  }
	\label{fig:nico}
\end{figure}

 \begin{figure*}[htb]
\vspace{-2.3cm}
	\hspace{-1.3cm}
		\includegraphics[width=1.15\textwidth]{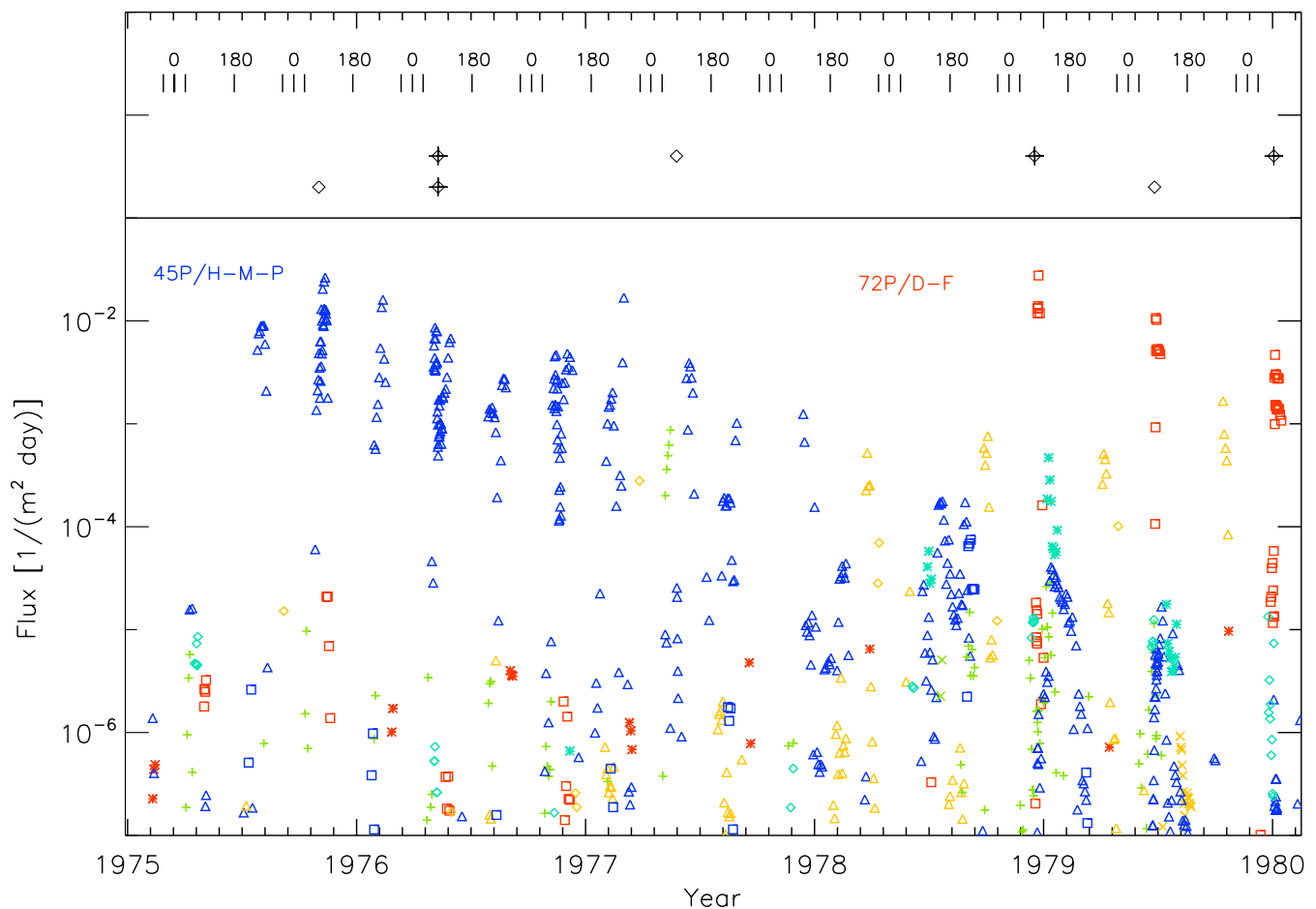}
		\vspace{-1.7cm}
	\caption{Simulated dust fluxes for cometary meteoroid trails intercepted by the Helios spacecraft (cf. Fig.~\ref{fig:orbitplot}). 
	Symbols and colours distinguish individual comets. Helios' true anomaly angle 
	is indicated at the top.
	Black diamonds show the detection times of  seven particles at a true anomaly angle of 
	$\eta=135\,\pm 1^{\circ}$, four particles identified in this work as potential cometary trail particles are additionally marked with 
	crosses (top row: detections with the ecliptic sensor; bottom row: detections with 
	the south sensor).  The colour symbols refer to the 
	following comets: red squares: \denning;  blue triangles: \honda; green crosses: \finlay; 
	light blue asterisks: \machholz; yellow triangles: 210P/Christensen. The remaining symbols refer to other comets forming a 
	very low background flux. 
    The simulations were performed with a two-day timestep, and the simulated particles are in the mass range 
	$\mathrm{10^{-8}\,kg} \leq m \leq \mathrm{10^{-2}\,kg}$. 
	  }
	\label{fig:flux}
\end{figure*}

\citet{altobelli2006} re-analysed the Helios dust data, searching for interstellar particle impacts
in the inner solar system. 
When analysing the data  as a function of Helios' true anomaly angle $\eta$, the authors recognized a cluster of 
seven impacts in a very narrow range  $\eta=135\,\pm 1^{\circ}$.  Figure~\ref{fig:nico} shows  a subset of the 
Helios dust data  together with these  cometary trail particle candidates.
 These data were 
 obtained during a total of ten Helios orbits around the Sun. 
  
 The particle concentration at $\eta=135^{\circ}$ 
is indicated by a vertical solid line. These seven impacts were
 detected during six Helios orbits in a very narrow spatial range  
 between 0.72 and 0.75~AU 
distance from the Sun (two impacts occurred on the same 
day, see  Table~\ref{tab:particles}). The derived particle masses were in the range 
$\mathrm{10^{-16}\, kg} \lesssim m \lesssim \mathrm{10^{-12}\,kg}$ (Table~\ref{tab:particles}), with an uncertainty of 
a factor of 10 in the mass calibration of a single particle. This 
remarkable coincidence of repetitive detections at approximately the same location led 
the authors to speculate that the impacts may have occurred when the Helios spacecraft repeatedly traversed the
meteoroid trail of a comet. The authors argued that owing to their 
size, such grains would be little sensitive to radiation pressure, and they would keep the orbital elements 
of their parent body for some time. 
The hypothesis, however, could
not be investigated further because no comprehensive dust trail model was available at the time. Here 
we study this hypothesis further.

\section{IMEX Cometary Trails Model}

\label{sec:model}

\begin{table*}[htbp]
   \caption{Orbital data of comets discussed in this paper from the JPL Small Bodies Database (ssd.jpl.nasa.gov) 
   if not stated otherwise, 
   as well as the simulated approximate particle impact speed $v_{\mathrm{imp}}$ at $\eta = 135^{\circ}$ 
   (column~9).}
   \centering
   \begin{tabular}{@{} lcccccccc @{}} 
         \hline
\multicolumn{1}{c}{Comet}               & $e$  &   $q$  &    $i$     & $\Omega$   & $\omega$   &  $t_{\mathrm{Perihelion}}$   & Epoch & $v_{\mathrm{imp}}$\\  
                                        &      &  [AU]  &[$^{\circ}$]&[$^{\circ}$]&[$^{\circ}$]&               &             &     [\kms]   \\
\multicolumn{1}{c}{(1)}                 &  (2) &   (3)  &    (4)     &     (5)    &     (6)    &       (7)     &     (8)     & (9) \\
   \hline
 45P/Honda-Mrkos-Pajdu\v{s}{\'a}kov{\'a}& 0.81 &  0.58  &  13.1      &    233.7   &      184.5 &   28-Dec-1974  & 19-Dec-1974       & 32 \\
  72P/Denning-Fujikawa                  & 0.82 &  0.78  &  9.2       &    36.1    &     337.9  &   02-Oct-1978$^{\ast}$& 20-Nov-2014& 21 \\	
      \finlay                           & 0.70 &  1.10  &  3.65      &    42.4    &     322.2  &   03-Jul-1974  & 12-Jul-1974       &  20 \\
    \machholz                           & 0.75 &  0.75  &  12.8      &   246.2    &     149.3  &   18-Sep-1994  & 05-Sep-1994       & 24 \\
      \hline
      $\ast$: \citet{sato2014}
   \end{tabular}
   \label{tab:comets}
\end{table*}

\begin{figure*}[tb]
	\vspace{-2.8cm}
	\hspace{-0.3cm}
		\includegraphics[width=0.53\textwidth]{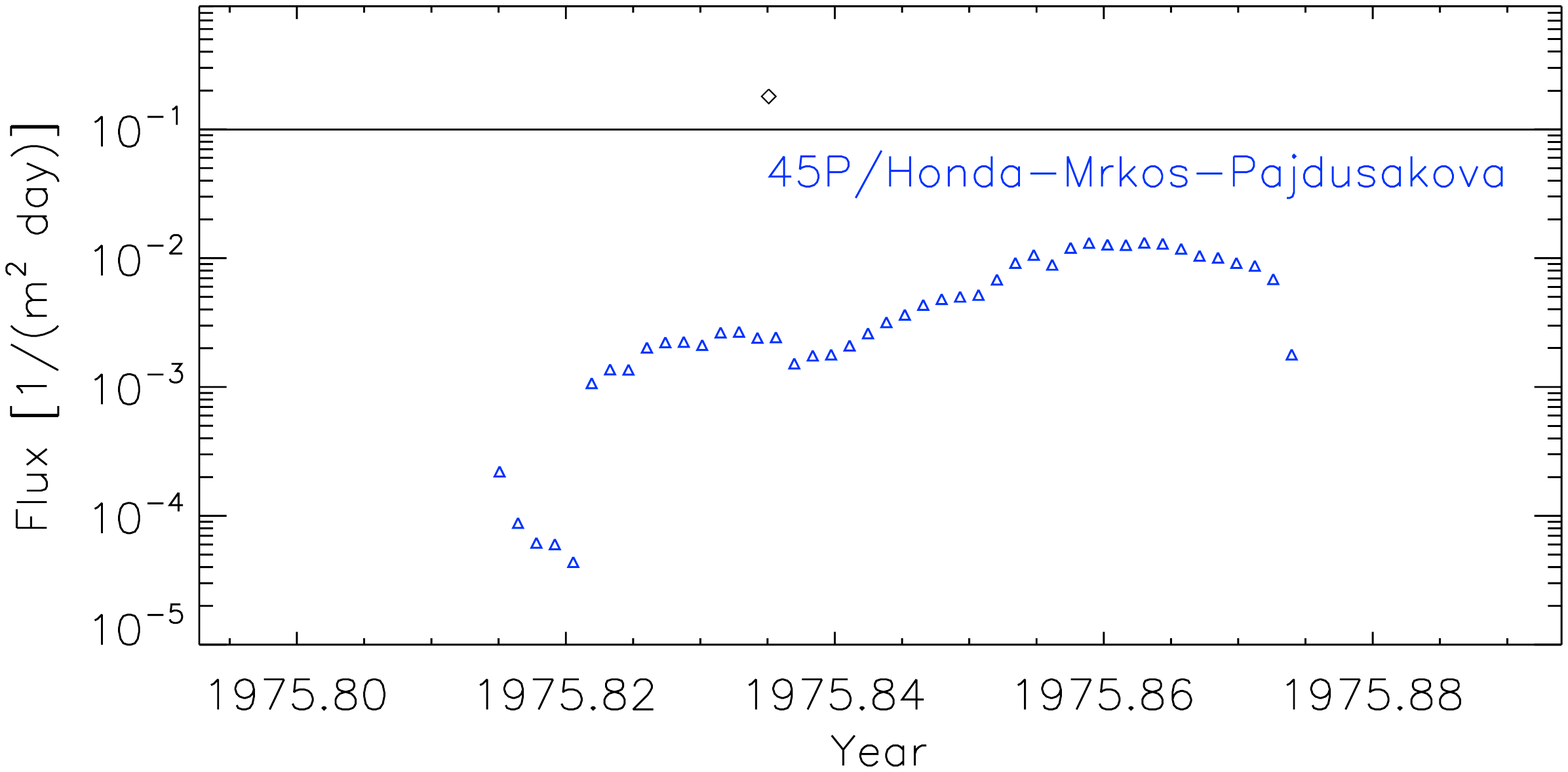}
		\hspace{-0.6cm}
		\includegraphics[width=0.53\textwidth]{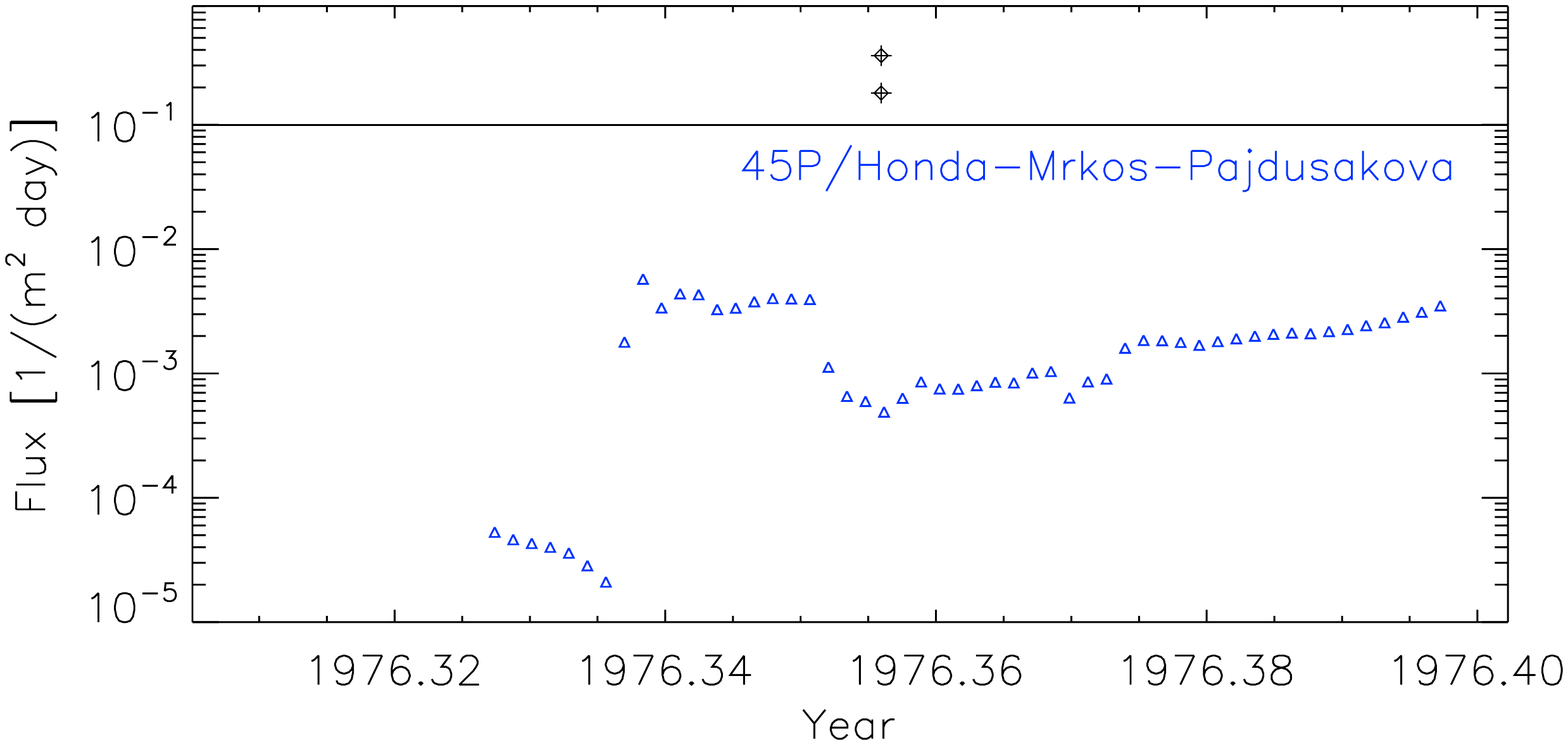}
		
		\vspace{-3.0cm}
		\hspace{-0.3cm}
		\includegraphics[width=0.53\textwidth]{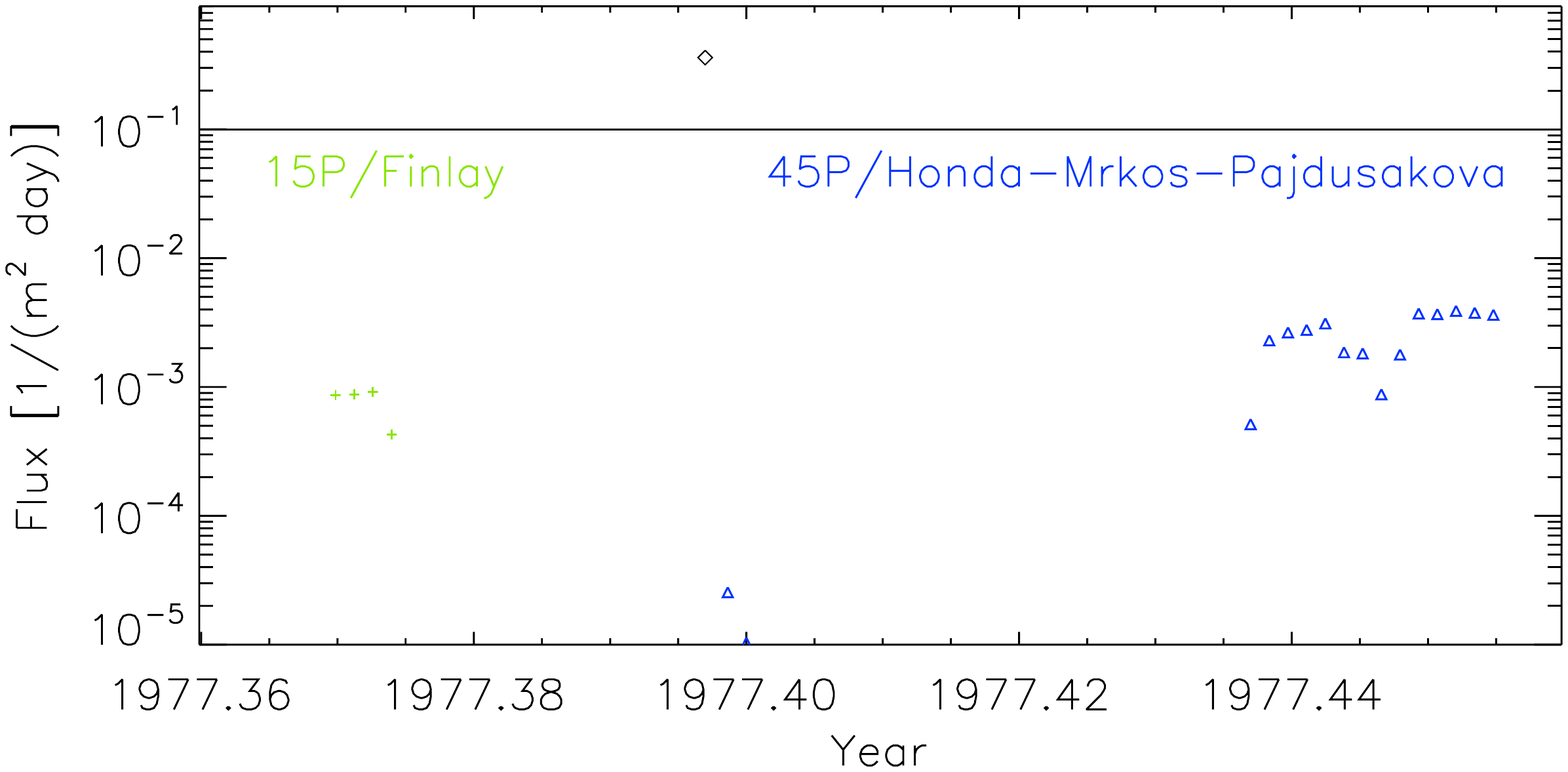}
		\hspace{-0.6cm}
		\includegraphics[width=0.53\textwidth]{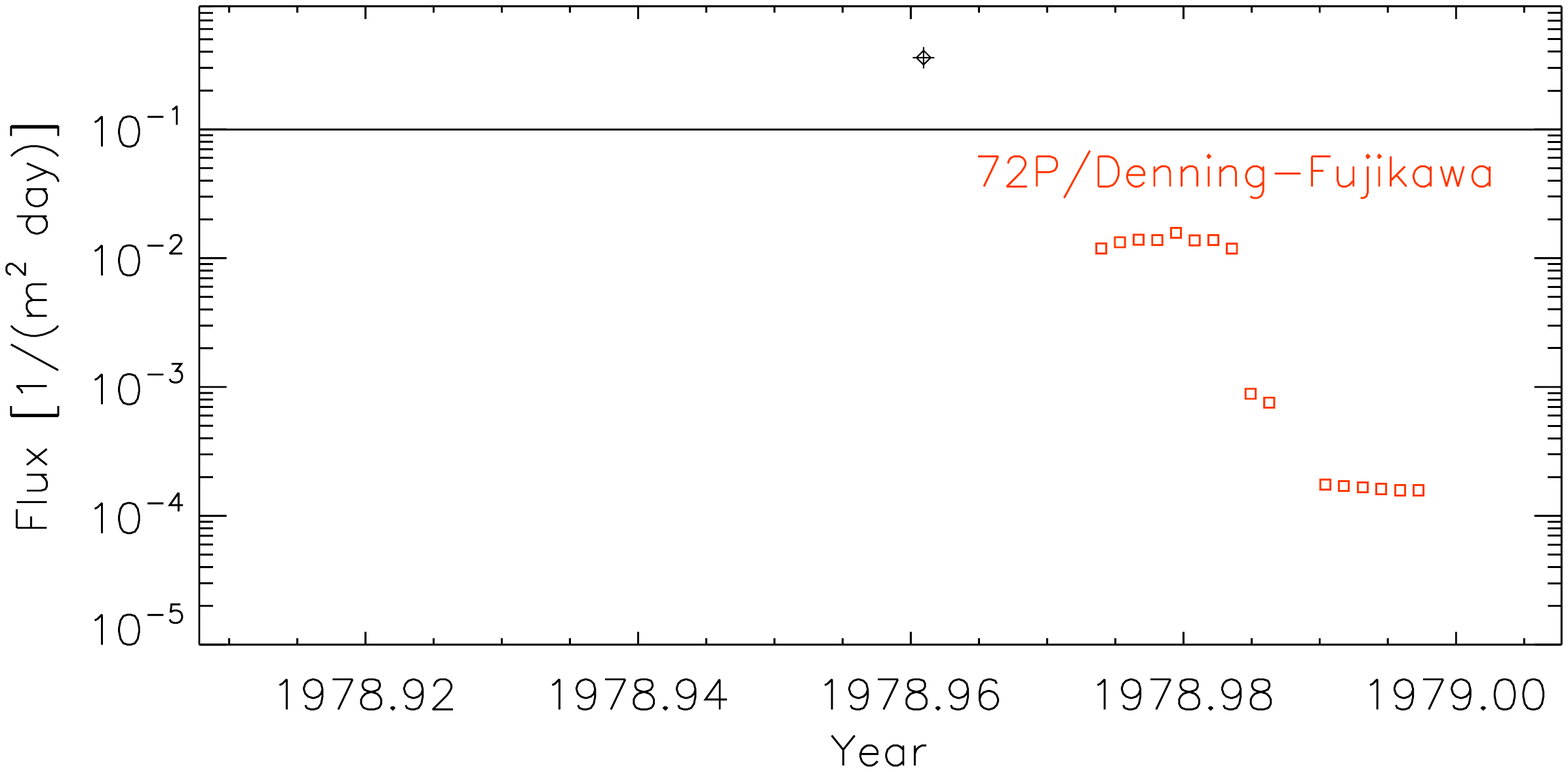}
		
				\vspace{-3.0cm}
		\hspace{-0.3cm}
		\includegraphics[width=0.53\textwidth]{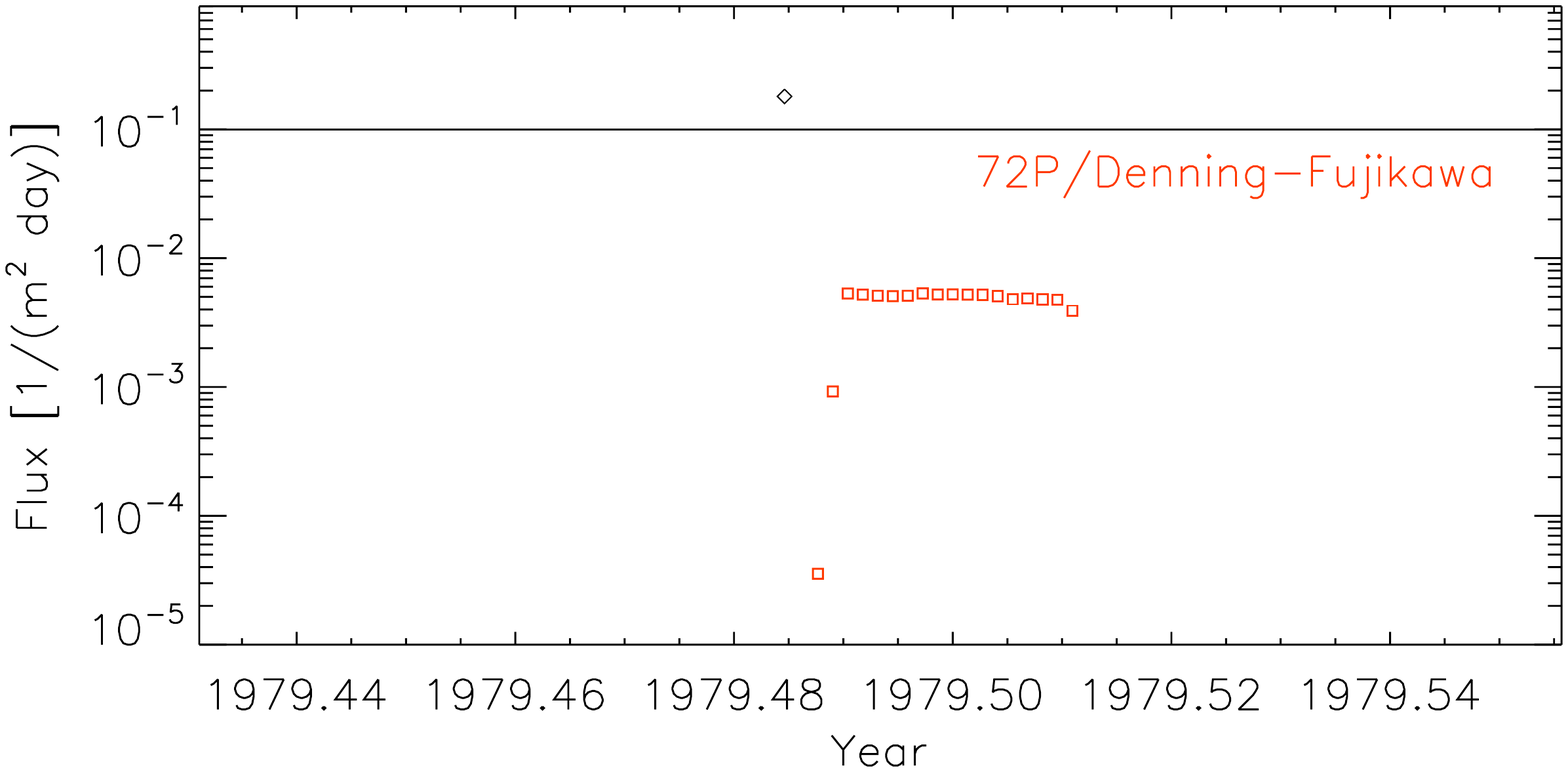}
		\hspace{-0.6cm}
		\includegraphics[width=0.53\textwidth]{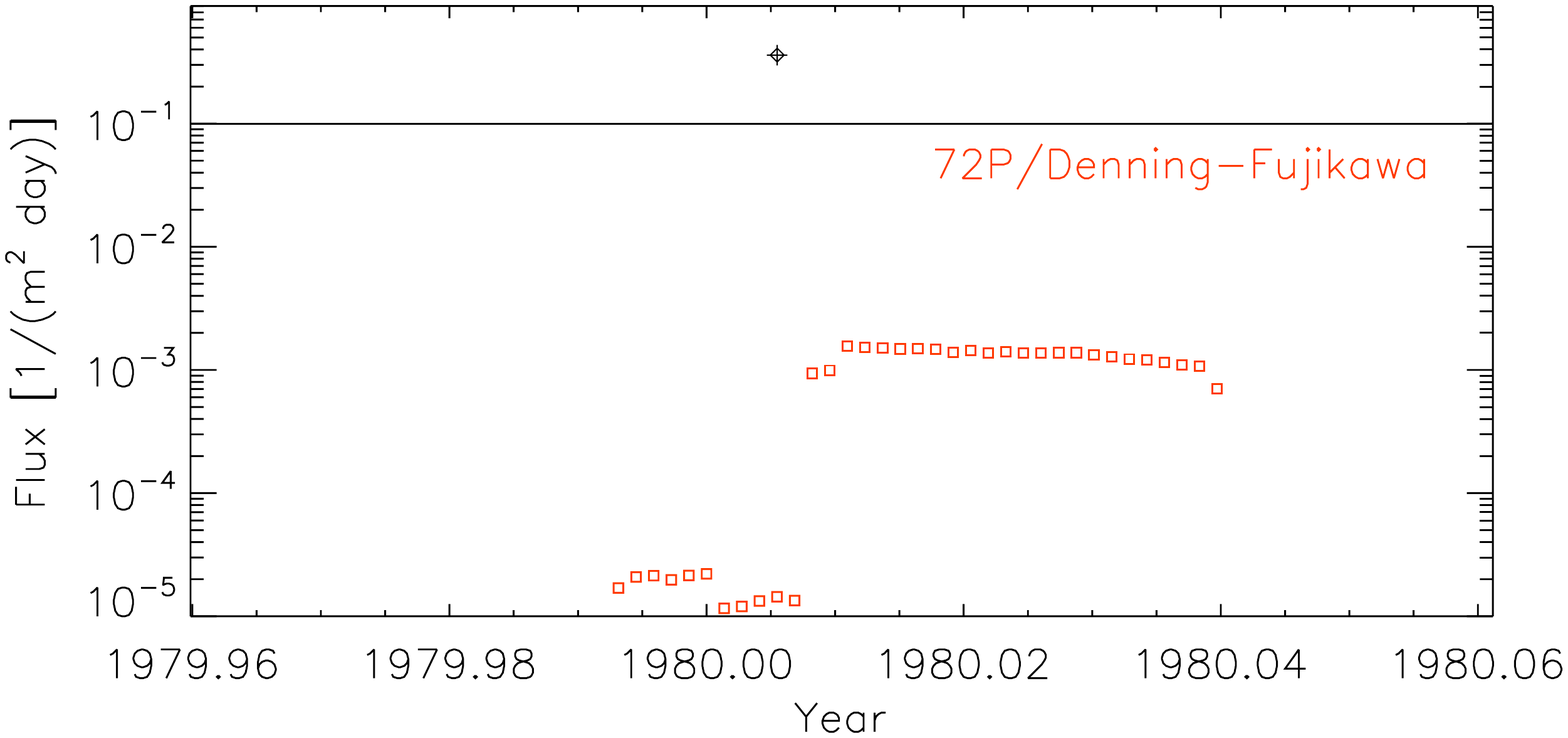}
		
		\vspace{-0.4cm}
	\caption{Same as Figure~\ref{fig:flux} but for the time periods when Helios detected the candidate cometary trail particles. 
	For each panel the time interval shown is one month and the separation of each simulated data point is 12~hours. The 
	 times of the Helios 
	detections are indicated  at the top.
	  }
	\label{fig:fluxhigh}
\end{figure*}

In order to identify time intervals when the Helios spacecraft traversed cometary meteoroid trails, we use
the Interplanetary Meteoroid Environment for eXploration (IMEX) dust streams in 
space model developed by \citet{soja2015a,soja2015b,soja2019}.  The model generates  trails for 362 Jupiter-family, 
40 Halley-type, and 18 Encke-type comets available in the JPL 
Small Body Database (SBDB) as of 1 August 2013, which have perihelion distances $q$ < 3 AU, semimajor axes $a$ < 30 AU and 
defined total visual magnitudes. 

Particles are emitted when the comet is in the inner solar system, taking into account 
comet perihelion passages between the years 1700 and 2080  for Encke-type comets, and between 1850 and 2080 for 
Jupiter-family and Halley-type comets, respectively. This reflects the fact that the most recent dust is expected 
to be most important, and also the maximum size of the database that could be maintained at the time when the model was developed. 

For each passage through the inner solar system  within 3~AU of the Sun of each comet (which we call apparition in the following),
particles are emitted randomly from the comet's sunlit hemisphere of the comet nucleus within the time ranges specified above. 
About 28000 particles are ejected per comet per apparition for Halley-type comets; and about 14000 
for other comets.

The dust ejection is described by the velocity model from the hydrodynamic comet emission model of 
 \citet{crifo1997}. The model assumes the dust emission to be driven by  water gas production within 3~AU
 distance from the Sun.

The model estimates the water production rate  
using the visual magnitude, and a gas-to-dust ratio based on an empirical formula given by \citet{jorda2008}.
The JPL Small Body Database provides total and nuclear magnitudes.

Dust-to-gas mass ratios can be estimated for individual comets, and they mostly range from 0.1 to 3, 
though higher values are possible. Furthermore, they appear to be dependent on heliocentric distance \citep{ahearn1995}. 
Given the large uncertainties in dust-to-gas ratios, the model uses a value of 1. 
Deviations from this can be considered in the analysis of individual comets.

The IMEX model uses the mass distribution model of \citet{divine1987} and \citet{agarwal2007a,agarwal2010}, with model parameters
given by \citet{soja2015a}. 
The  mass distribution covers the  range from 
$\mathrm{10^{-8}\,kg}$ to $\mathrm{10^{-2}\,kg}$, separated into eight mass bins \citep[approximately corresponding to 
$\mathrm{100\, \mu m}$ to 1~cm particle radius;][]{soja2015a}. 
The particle density is assumed to be $\mathrm{\rho = 1000\,kg\,m^{-3}}$. For comets with  unknown radius a value of 
1~km is assumed \citep{soja2015a}.

The trajectory of each emitted particle is integrated individually including solar gravity, 
planetary perturbations as well as solar radiation pressure and Poynting-Robertson drag. 
 Non-gravitational cometary forces are neglected because they are not well known for most comets, and their
effect is largely to alter the location of the comet in its orbit, rather than the orbit itself.

Due to storage space considerations, the particle state vectors were saved only during a limited time interval from 1980 to 2100. In order to compare the IMEX Streams model to the Helios data from 1975 to 1980, we identified candidate comets by extrapolating the simulated particle state vectors backwards from 1980 using only solar gravity and radiation pressure, subsequently re-doing the full integration for the two comets with the highest flux, i.e. 45P/Honda-Mrkos-Pajdu\v{s}{\'a}kov{\'a} (hereafter \honda), and
72P/Denning-Fujikawa (hereafter \denning), and storing their particle state vectors starting from 1960.

The
model calculates the impact velocity for each individual particle on to the 
spacecraft as well as dust number density and flux. We use the IMEX model to identify time intervals when Helios traversed the 
meteoroid trails of comets between December 1974 and January 1980 when 
dust measurements are available.  A detailed model description including an 
application to the  trail of comet 67P/Churyumov-Gerasimenko was given by \citet{soja2015a}. 

\section{Results}

\label{sec:results}

In this Section we present the results of our dust trail simulations for the time period 
between 19 December 1974 and 02 January 1980 when 
Helios collected dust measurements. In this time interval 
the spacecraft completed ten revolutions around the Sun and repeatedly 
traversed the meteoroid trails of several comets.
We compare the times when Helios detected the particles and the measured impact directions with
the model predictions in order to constrain the particle sources.

\subsection{Simulated Dust Fluxes}

\label{subsec:fluxes}

In Figure~\ref{fig:flux} we show the simulated fluxes for Helios' cometary trail traverses. The simulations 
 identified the trails of 13 comets that were traversed by Helios. The predicted fluxes for most of these 
crossings are below approximately $\mathrm{10^{-4}\,m^{-2}\,day^{-1}}$, which is insignificant 
for our analysis (for some comets the predicted flux is even below $\mathrm{10^{-7}\,m^{-2}\,day^{-1}}$ and therefore
not shown in the diagram). The maximum dust fluxes 
predicted for  trail traverses of individual comets  
vary by up to four orders of magnitude. 

\begin{figure}[tb]
	\vspace{-0.4cm}
	\hspace{-1.2cm}
		\includegraphics[width=0.66\textwidth]{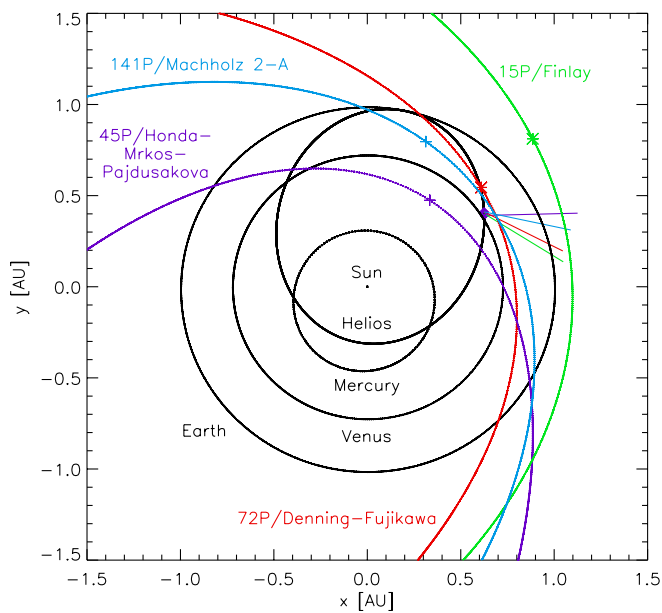}
		\vspace{-7.4cm}
	\caption{Orbits of Helios and the candidate comets. The purple diamond marks the dust detections 
	 at $\eta \approx 135^{\circ}$. The  lines attached to the 
	diamond indicate the approximate impact 
	directions (speed vector) of trail particles from these comets in the spacecraft-centric reference frame as derived from the IMEX model. 
	The X-Y plane is the 
	ecliptic plane with vernal equinox oriented towards the +X direction. Comet orbits are shown for
	the period 1975 to 1980, with locations of ascending nodes (asterisks) and descending nodes (plus signs) superimposed.
	  }
	\label{fig:orbitplot}
\end{figure}

A  repetitive pattern is obvious for comets \denningg (red squares) 
and \hondaa  (blue triangles):
Strong peaks  
occur during consecutive revolutions of Helios around the Sun. The model predicts maximum fluxes of approximately 
$\mathrm{3 \cdot 10^{-2}\,m^{-2}\,day^{-1}}$ for these two comets.
The flux peaks are rather narrow with a typical peak width of approximately 5 to 20 ~days. Both are Jupiter 
family comets with orbital periods of 5 and 9~years, respectively (Table~\ref{tab:comets}). Interestingly, for 
both comets 
the flux predicted for each trail traverse decreases with time for consecutive traverses (Figure~\ref{fig:flux}).
Helios' trail traverses occurred close after these two comets passed through their perihelia (Figure~\ref{fig:orbitplot} 
and Table~\ref{tab:comets}), and this decreasing flux is in agreement with a drop in the dust density along 
the trail with increasing distance from the comet nucleus. 

At the top of Figure~\ref{fig:flux} we indicate the detection times of the seven dust particles at a true
anomaly angle of 	$\eta=135\,\pm 1^{\circ}$ which \citet{altobelli2006}  recognized as candidate 
trail particles. The simulations show that close to all these detections 
Helios traversed at least one meteoroid  trail. 
This is particularly evident for comet \denning: A flux exceeding 
$\mathrm{6 \cdot 10^{-3}\,m^{-2}\,day^{-1}}$ is predicted for three trail traverses in December 1978, 
 June 1979 and  January 1980. Helios  detected one particle impact during or close to 
each of these traverses. 

There are also particle detections at the trail traverses of comet \hondaa in the 
interval 1975 to 1977. Two particles were detected during the traverse of this comet's trail 
in May 1976. On the other hand, there are traverses of the trail of this comet in 1976 
and 1977 at a true anomaly angle  $\eta \approx 225^{\circ}$ where there is no
obvious particle concentration in the Helios data set (cf. Figure~\ref {fig:nico}). However, at this location the
model predicts somewhat lower fluxes, making particle detections less likely, although the
detection of single unidentified trail particles by Helios cannot be excluded. 
Figure~\ref{fig:flux} also shows  that the 
 seven candidate  trail particles were particularly detected during trail traverses with the highest 
predicted dust fluxes. 

\begin{figure*}[tb]
	\vspace{-2.9cm}
	\hspace{-1.3cm}
		\includegraphics[width=1.15\textwidth]{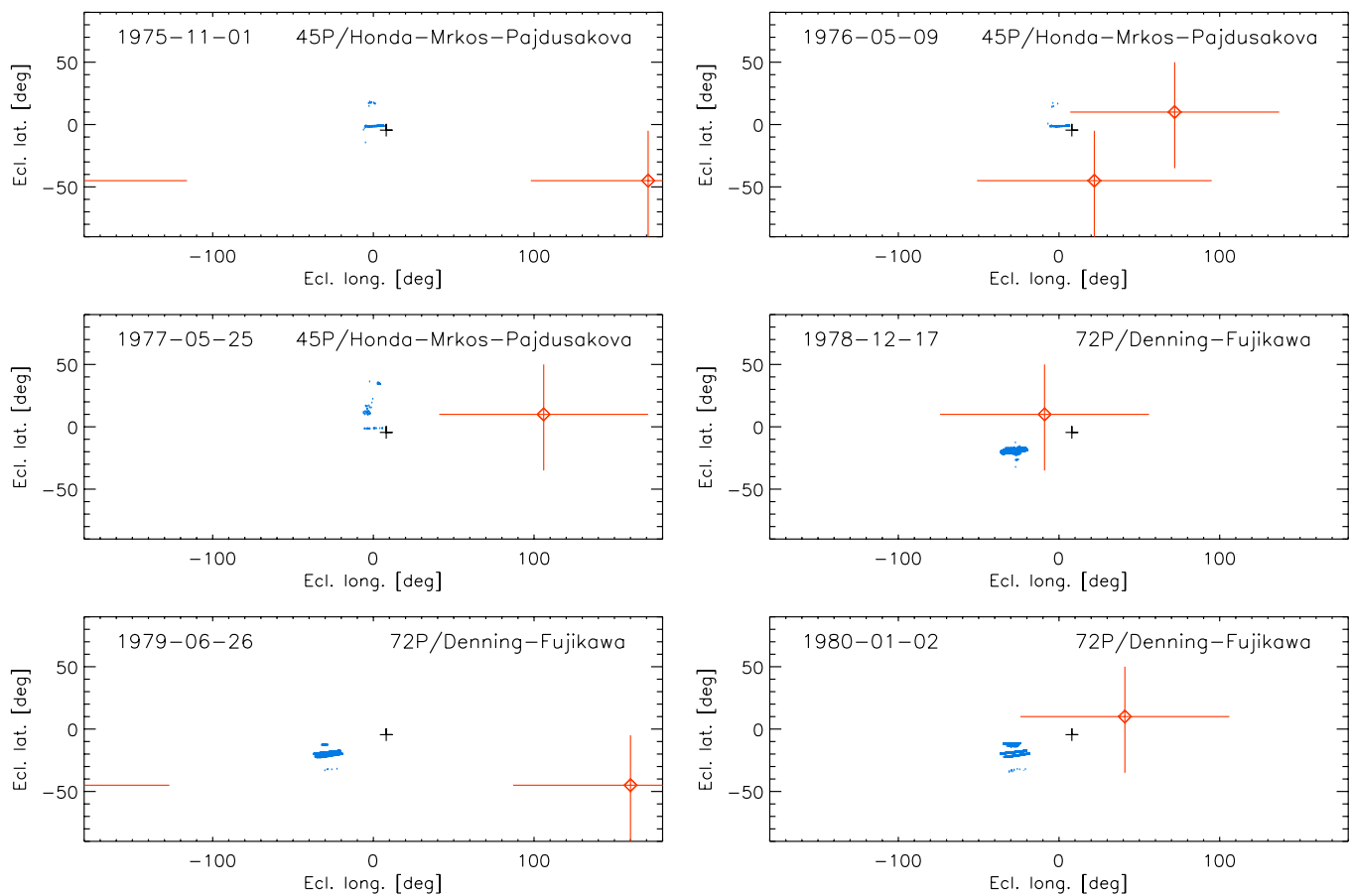}
		\vspace{-1.6cm}
	\caption{Impact directions of detected particles compared with the directions of simulated  trails
	for comets \denningg and \hondaa
	in the spacecraft-centric reference frame in ecliptic coordinates. Blue dots show the approach directions of
    simulated trail particles. Red diamonds indicate the sensor orientation during the time of 
	particle impact; the ecliptic longitude of the data point is the spacecraft spin
	orientation at the time of particle impact while the ecliptic latitude  corresponds
	to the maximum of the sensor sensitivity profile in  latitudinal direction \citep{gruen1980a}. 
	Red crosses indicate the approximate sensor field-of-view  (cf. Figure~\ref{fig:sensareas}), 
	and small black crosses show the impact direction of 
	particles orbiting the Sun on circular orbits at Venus' heliocentric distance.  The detection time of the impact 
	is given at the top left of each panel.
	  }
	\label{fig:directions}
\end{figure*}

In Figure~\ref{fig:fluxhigh} we compare in more detail the detection times of the candidate trail particles with 
the time intervals when the model predicts the highest  fluxes for comets \hondaa and \denning. 
Three  detections in 1975 and 1976 nicely coincide with the time interval when the 
model predicts  relatively high fluxes in the range of $\mathrm{10^{-2\ldots -3}\,m^{-2}\,day^{-1}}$ (top panels).
The two  detections in 1979 and 1980 (bottom panels) are offset from the highest predicted trail fluxes by only one day. 
In two cases, however, the offset is 4~days (1978) and 8~days (1977), respectively (middle panels). We will
have a more detailed look at the particle detection times in Section~\ref{sec:discussion}.

\subsection{Detection Geometry and Impact Speeds}

In addition to the detection time of the particles, the impact direction is another important parameter to 
constrain the particles' origin. 
In Figure~\ref{fig:orbitplot} we show the Helios trajectory together with  orbital sections 
for the 
comets that exhibit the highest meteoroid fluxes during trail traverses as 
shown in  Figure~\ref{fig:flux}. The simulated impact directions 
of particles on to the spacecraft in the spacecraft-centric reference frame are indicated at a true anomaly angle $\eta=135^{\circ}$,   
 corresponding to the trail traverses for which the model predicts the highest dust fluxes  
 (cf. Figure~\ref{fig:flux}). Orbital elements 
for these comets 
are listed in Table~\ref{tab:comets}. 

\begin{figure*}[tb]
	\vspace{-1.8cm}
	\hspace{-0.1cm}
		\includegraphics[width=0.69\textwidth]{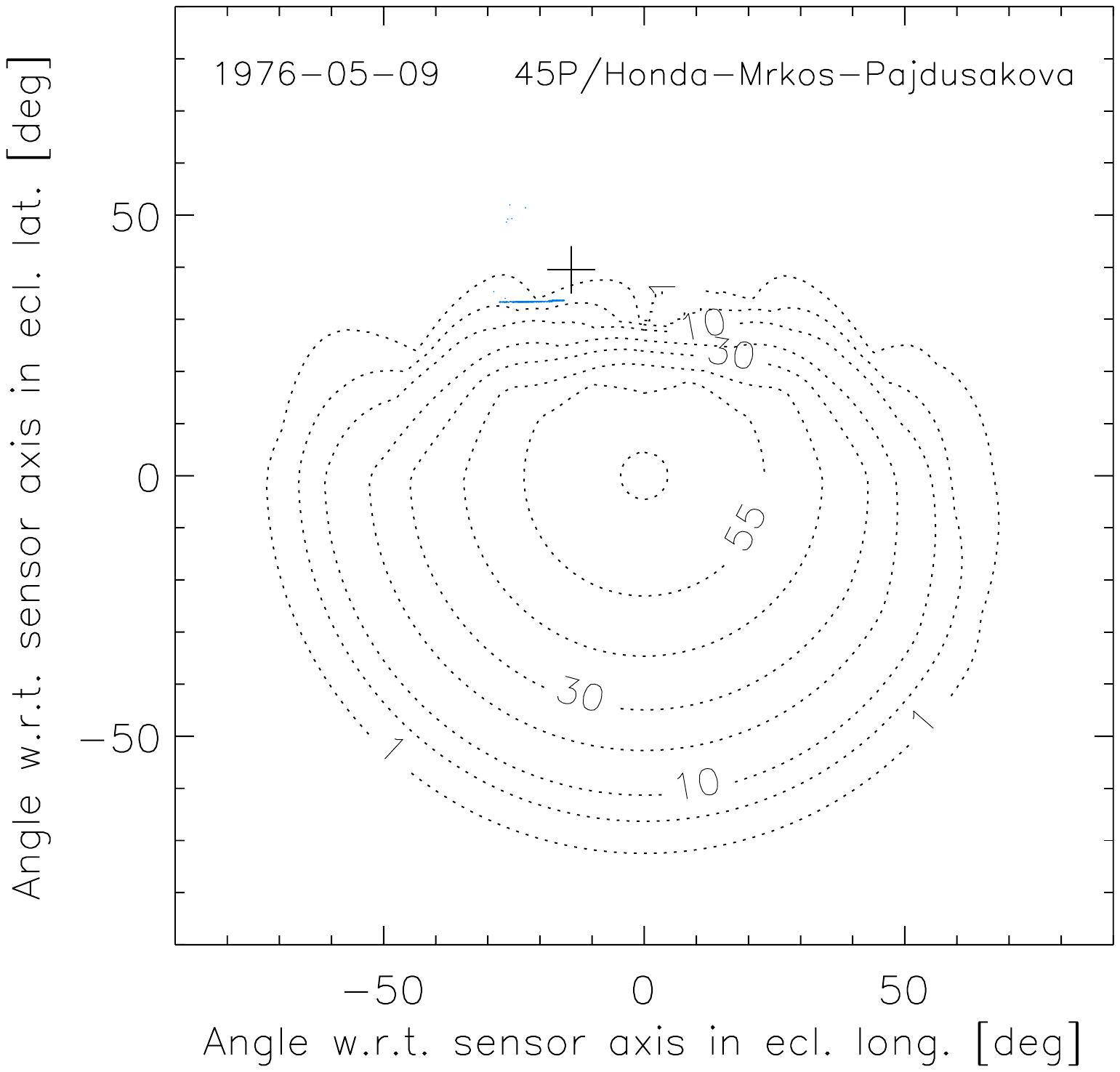}
		\hspace{-3.4cm}
		\includegraphics[width=0.69\textwidth]{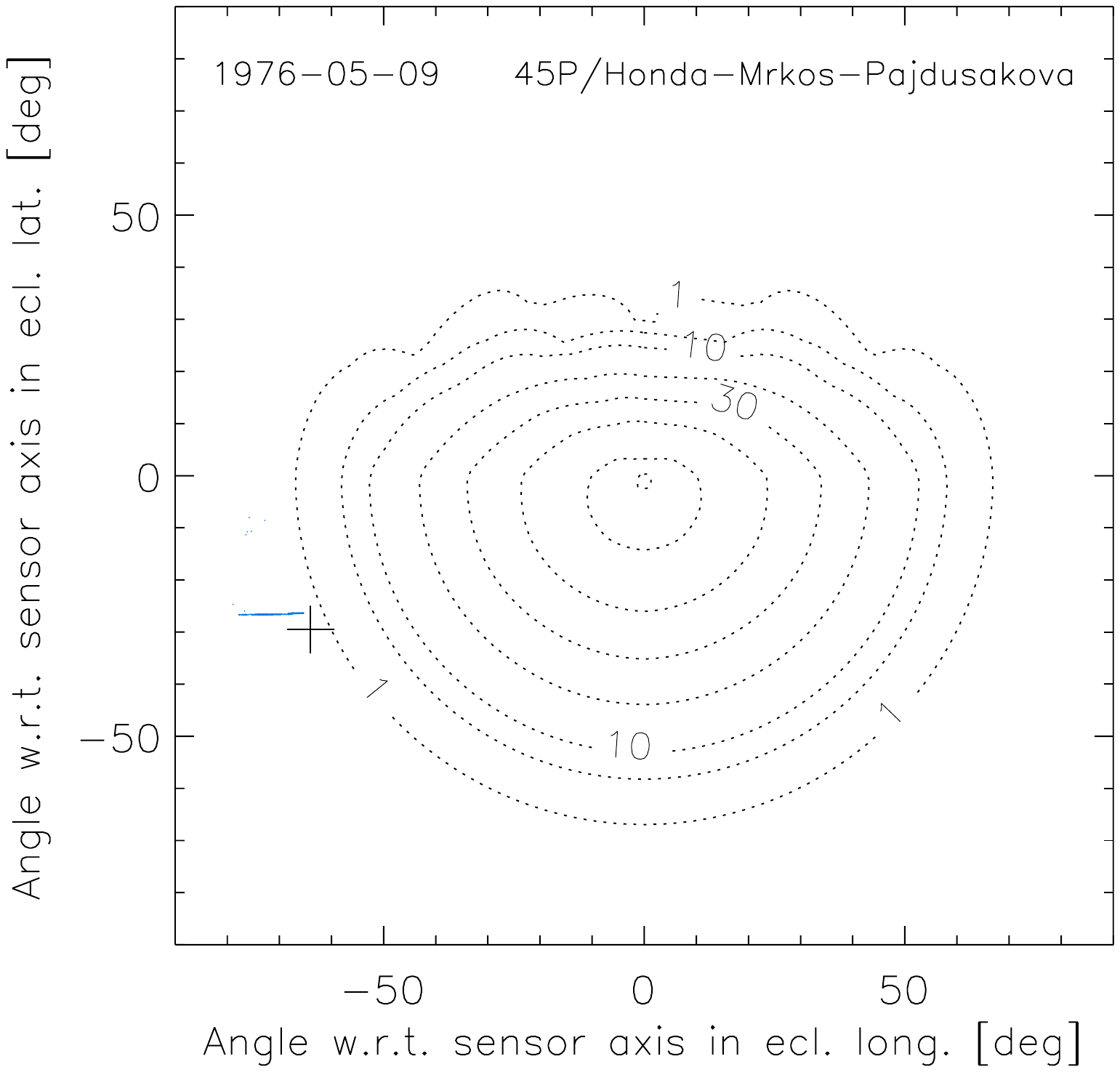}
		
		\vspace{-2.5cm}
		\hspace{-0.1cm}
		\includegraphics[width=0.69\textwidth]{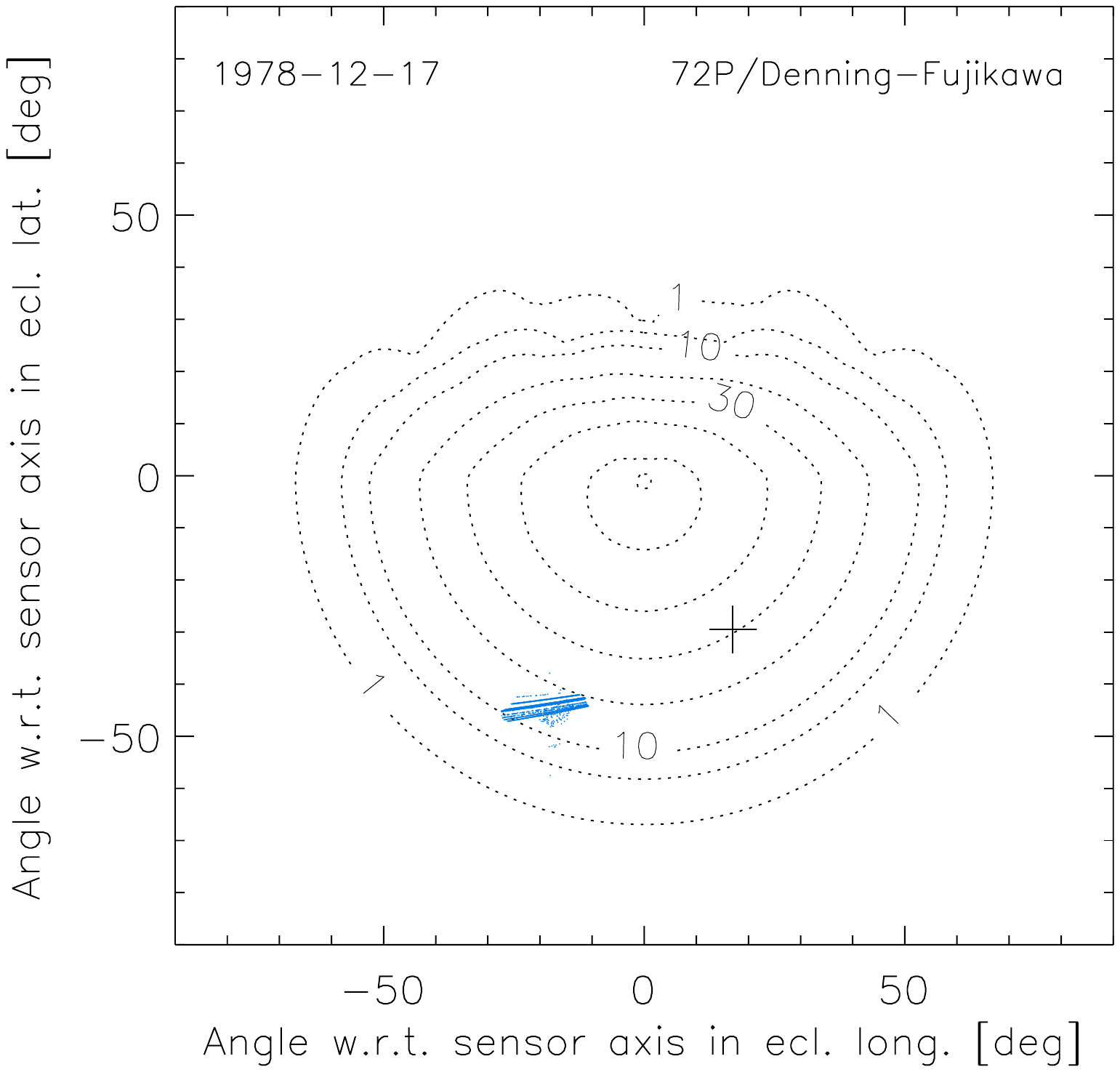}
		\hspace{-3.4cm}
		\includegraphics[width=0.69\textwidth]{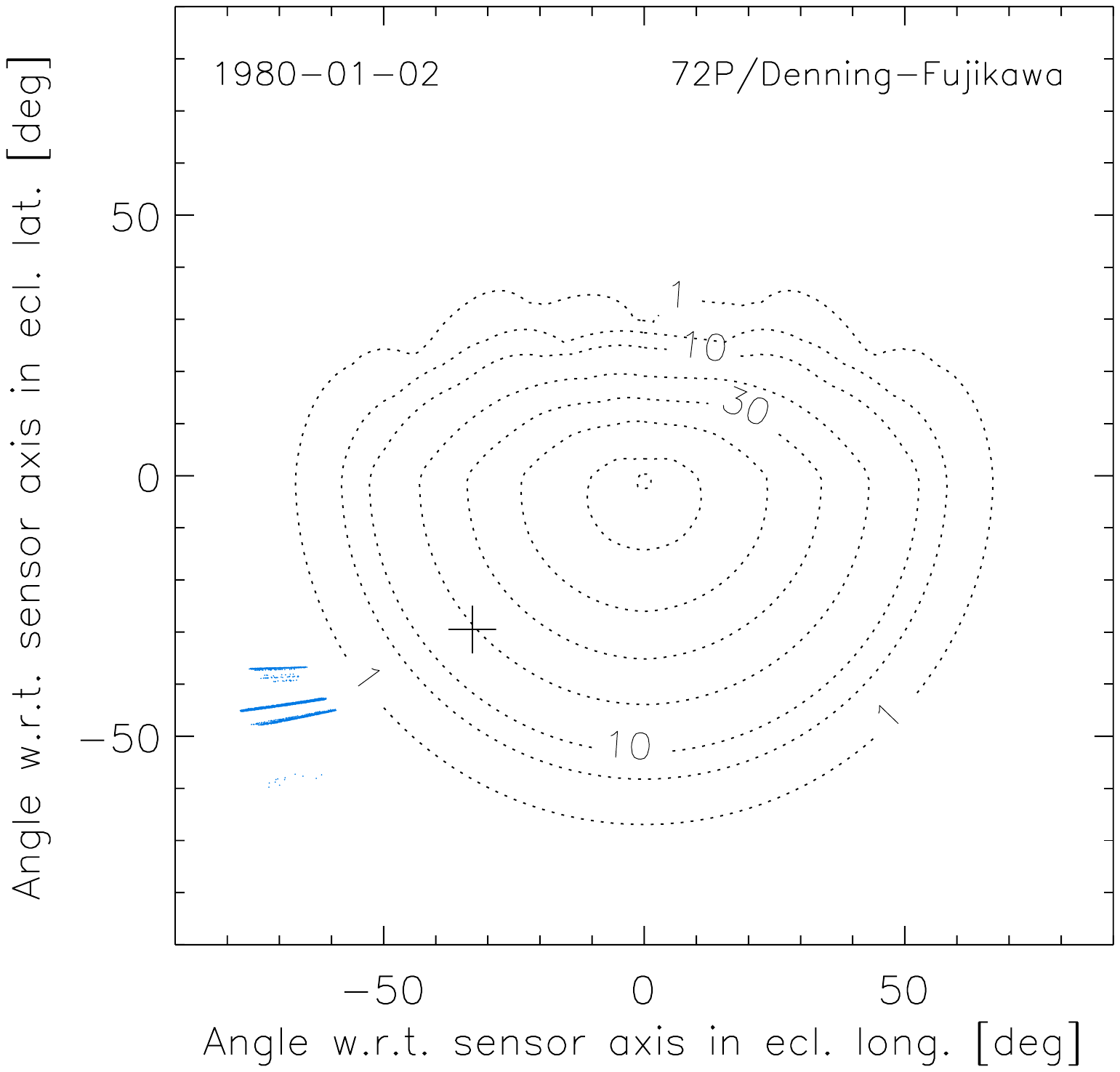}
		
		\vspace{-0.4cm}
	\caption{Sensor field-of-view for the south sensor (top left panel) and the ecliptic sensor (all other panels) from
	\citet{gruen1980a}.  The centre (coordinates [0,0]) is the direction of the sensor boresight. 
	The numbers at the 
	contour lines give the  sensitive area of the sensor target in square centimeters, which depends on the particle impact direction. 
	The simulated impact 
	directions of the cometary trail particles (blue dots), 
	and particles orbiting the Sun on circular orbits at Venus' orbit (black cross) are superimposed for the sensor pointing 
	at the impact time of the dust particle, see also Figure~\ref{fig:directions}. 
	  }
	\label{fig:sensareas}
\end{figure*}

For the relevant comets the simulated particle impact speeds are between  20\,\kms\ and 32\,\kms\ 
(Table~\ref{tab:comets}). The measured particle impact speeds are 
in the range of approximately 10 to 40\,\kms\ (Table~\ref{tab:particles}), which is in rather good agreement with these values, 
given that the single speed measurement has an uncertainty  of at least a factor of two.

Figure~\ref{fig:orbitplot} shows that by a remarkable coincidence, at a true anomaly angle of approximately $\eta=135^{\circ}$,
Helios traversed the trails of three comets: \honda, \denningg and \machholz, with the 
orbit of a fourth one, \finlay,  also being close. Furthermore, the traverse of 
Venus' orbit occurred in the same region. This coincidence suggests that the 
candidate trail particles could be particularly easily recognised as such because of this
concentration of trail traverses within a rather small region of space. 

From the spacecraft spin orientation at the time of dust detection we can constrain the
impact direction of each detected particle.  In Figure~\ref{fig:directions}  we compare  
the Helios detections with the simulated impact directions 
for trail particles released from comets \denningg and \honda. Here the largest uncertainties 
arise from the rather large sensor field-of-view (indicated by red crosses) while the spacecraft orientation 
is known with a high accuracy of better than $1.4^{\circ}$.
Some  
fine structure is evident in the simulated trails. The geometries for  \machholzz and \finlayy are 
very similar, however, they are not considered further because for them the  model predicts significantly
lower  fluxes than for the other two comets (cf. Figure~\ref{fig:flux}). 

In Table~\ref{tab:particles} we summarise our results for the particle detections. The
strongest criterion is the impact direction. If this is not compatible with a trail origin
we discard the particle from the list of potential trail particles. Figure~\ref{fig:directions} shows that 
this is the case for the
three detections in 1975, 1977 and 1979. From the remaining four particles, the particle 
measured in 
1978 is  compatible with an origin from comet \denning, while the three detections 
in 1976 and 1980 are marginally compatible with an origin from comets \denningg and \honda, respectively. The impact
times agree with this interpretation for the two detections in 1976, while there is an offset for 
the detections in 1978 and 1980. 
The impact speed is only listed in Table~\ref{tab:particles} for comparison, it is not 
used as a criterion for trail identification. 

The  fields-of-view of the Helios dust sensors are shown in Figure~\ref{fig:sensareas}. For 
the dust impacts measured in 1976, 1978, and 1980, which are in best
agreement with a cometary trail origin, we also show the impact directions of the simulated  trails of comets \hondaa and \denning, as well as 
the approximate directions of Venus dust ring particles orbiting the Sun on  circular heliocentric orbits. As was 
already concluded from  Figure~\ref{fig:directions}, the 
particle detected in 1978 was well within the dust sensor field-of-view while the three other detections were likely  close to the
edge of the field-of-view. The sensor side wall is not taken into account in Figure~\ref{fig:sensareas}  (cf. Section~\ref{sec:discussion}).

It should be emphasised that for our analysis we have only used directly measured parameters, 
like the sensor azimuth and the 
spacecraft true anomaly angle at the time of particle impact. We only refer to derived 
physical parameters like  impact 
speed and mass  to check for consistency with our simulation
results.  Therefore, our analysis is free of any 
uncertainties of the type introduced by empirical calibrations applied  to derive these
physical parameters from the measured quantities.                                                                                                                                                                                                                            

\subsection{Mass Spectra}
\label{sec:massspectra}

\begin{figure}[tb]
	\vspace{-0.5cm}
	\hspace{-1.0cm}
		\includegraphics[width=0.64\textwidth]{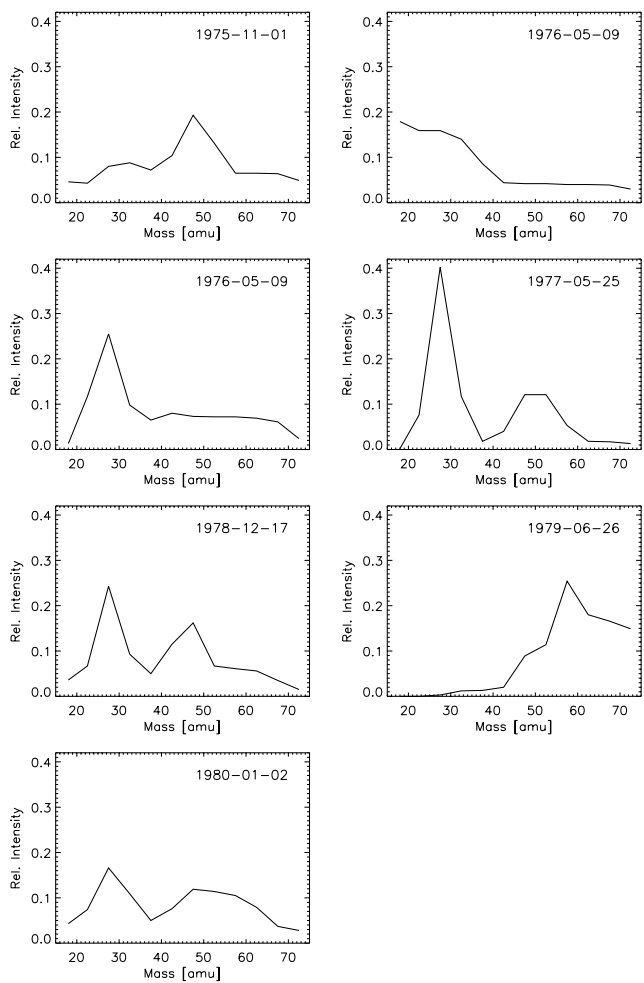}
		\vspace{-1.8cm}
	\caption{Mass spectra of the seven dust particles detected at a true anomaly angle of 
	$\eta=135\,\pm 1^{\circ}$. The day when the impact occurred is given at the top right
	of each panel.
	  }
	\label{fig:spectra}
\end{figure}

In addition to impact speed and particle mass, the Helios dust instruments  measured 
the particle composition with low mass resolution. The mass spectra of the 
seven candidate cometary trail particles are shown in Figure~\ref{fig:spectra}. We will
discuss them in more detail in Section~\ref{sec:discussion}.

\section{Estimation of Dust Fluxes from the  Measurements}

\label{sec:flux}

Our IMEX trail simulations show that up to four of the seven candidate cometary
trail particles detected by Helios are compatible with an origin from comet \hondaa
or \denning, respectively. Based on these trail identifications we attempt to constrain the dust fluxes 
in these trails by combining measurements and model results.   

The case of a single particle detection with an in-situ dust detector was considered by 
\citet{hirn2016}. The authors applied Poisson statistics to the measurements performed by the 
Dust Impact Monitor (DIM) on board the Rosetta lander Philae at comet 
67P/Churyumov-Gerasimenko. DIM detected a single particle impact during Philae's descent to the comet 
surface \citep{krueger2015b}.
Here we apply a similar approach to our Helios detections.

We assume that the cometary trail is a closely collimated stream of particles and that the 
impacts on to the Helios sensors are independent events, hence they should 
follow a Poisson distribution. For the periods when exactly one 
impact was detected during a trail traverse, only an upper limit for the ambient trail flux can be estimated. 
We  define the upper limit of the expected number of impacts as the highest value of $\lambda$ 
for which there is an arbitrarily chosen 5\% probability that the number of  detected events 
$N$ is less than two in a single measurement:
\begin{eqnarray} \label{eq:P}
P(N <2) & = & P(N =0)+P(N =1) \\    \nonumber
        & = & \frac{\lambda^0 \exp(-\lambda)}{0!} + \frac{\lambda^1 \exp(-\lambda)}{1!} \\
        & = & (1 + \lambda) \exp(-\lambda) = 0.05 \nonumber
\end{eqnarray}
resulting in 
\begin{equation}  \label{eq:lambda}
\lambda \approx 4.74. 
\end{equation}
The maximum impact rate is 
\begin{equation}
N_{\mathrm{max}} = \lambda_{\mathrm{max}}/T_{\mathrm{meas}},   
\end{equation}
where $\lambda_{\mathrm{max}}$ is given by Equation~\ref{eq:lambda} and $T_{\mathrm{meas}}$ is
the measurement time. For $T_{\mathrm{meas}}$  we assume the duration of a trail traverse predicted
by the model  
which is typically 10~days  (cf. Section~\ref{subsec:fluxes}).
The maximum flux on to the sensors is given by 
\begin{equation}
\Phi_{\mathrm{max}} = \frac{N_{\mathrm{max}}}{A} = \frac{\lambda_{\mathrm{max}}}{T_{\mathrm{meas}} \,\, A} \label{eq:phi},
\end{equation}
where A is the spin-averaged effective sensor area. 

Finally, the dust spatial density $D$ is given by 
\begin{equation} 
D_{\mathrm{max}} = \frac{\Phi_{\mathrm{max}}}{v_{\mathrm{imp}}}, \label{eq:density}
\end{equation}
where $v_{\mathrm{imp}}$ is the impact speed of the particles. 

\citet{gruen1980a} give 
sensor areas of $\mathrm{54.5\,cm^2}$ for the ecliptic sensor, and $\mathrm{66.5\,cm^2}$ 
for the south sensor,
respectively. Given that both sensors were always operated simultaneously, we simply add 
the two areas to obtain a total area of $\mathrm{121\,cm^2}$ for both sensors together. Due to the spacecraft 
spin, the spin-averaged effective sensor area was about a factor of four smaller, i.e. 
$A\approx\mathrm{30\,cm^2}$. With these numbers Equation~\ref{eq:phi} gives an upper limit for 
the dust flux in the trail of comet \denning:
\begin{equation}
\Phi_{\mathrm{max,72P}} = 158\,\mathrm{m^{-2}\,day^{-1}} \label{eq:phi_1},
\end{equation}  
and with the impact speed derived from the model given in Table~\ref{tab:particles}, the upper limit for the dust spatial density becomes:
\begin{equation}
D_{\mathrm{max,72P}} =  9\cdot 10^{-8}\,\mathrm{m^{-3}} \label{eq:d_1}.
\end{equation}  
These upper  limits apply to the cases when Helios detected one single particle per trail traverse. 

\begin{table*}[htbp]
   \caption{Data of the candidate cometary trail particles considered in this work. Detection day, 
   sensor which detected the particle, measured impact speed, measured particle mass, impact speed derived from  IMEX simulation, 
   compatibility with  trail origin based on given parameters, 
   source comet, flux in trail from source comet. Measured data in columns~1 to 4 are from \citet{gruen1981b}.}
   \centering
   \tiny
   \begin{tabular}{@{} lccccccccccc @{}} 
         \hline
\multicolumn{1}{c}{Day}           &  Sensor  & Measured  &     Measured                     &  \multicolumn{3}{c}{Compatibility with Trail Origin} &  Source   & Simulated   &  Trail Flux  & Dust Density\\  
                                  &          & Impact Speed &  Particle Mass                &   \multicolumn{3}{c}{ based on}                       & Comet     & Impact Speed & $\Phi$ & $D$ \\
                                  &          &  [\kms] &      [kg]                          &    Time &  Direction &  Speed                        &       &       [\kms]    & [$\mathrm{m^{-2}\,day^{-1}}$] & [$\mathrm{m^{-3}}$]\\
\multicolumn{1}{c}{(1)}           &   (2)    &   (3)   &      (4)                           &  (5)          &   (6)        &   (7)                 &   (8) & (9) & (10)  & (11) \\[0.5ex]
   \hline
                                  &                    &                                    &               &              &   &               &              & & \\[-2ex]
1975-11-01  &  S                  & $100^{+100}_{-50}$ &  $5.1^{+46}_{-4.6}\cdot 10^{-20}$  &  Yes          &   No         & -- &  --          &           -- &  -- & -- \\[0.5ex]
1976-05-09  &  S                  &  $11^{+8}_{-5}$    &  $5.4^{+375}_{-4.6}\cdot 10^{-13}$ &  Yes          &Possibly&(No)& \raisebox{-2.5mm}[0mm][2mm]{\honda}&\raisebox{-2.5mm}[0mm][2mm]{$33\pm 1.8$} & \raisebox{-2.5mm}[0mm][2mm]{$\approx 67$}  & \raisebox{-2.5mm}[0mm][2mm]{$\approx 2\cdot 10^{-8}$}\\[0.5ex]
1976-05-09  &  E                  &  $34^{+26}_{-15}$  &  $7.1^{+493}_{-6.1}\cdot 10^{-15}$ &  Yes          &   Possibly     & (Yes) &          &  \\ [0.5ex]
1977-05-25  &  E                  & $18^{+14}_{-8}$    &  $1.5^{+8.5}_{-1.3}\cdot 10^{-15}$ &  Possibly     &   No         & -- &    --          &        --    & -- & -- \\	[0.5ex]
1978-12-17  &  E                  & $9^{+6}_{-4}$      &  $2.2^{+13}_{-1.9}\cdot 10^{-16}$  &  Possibly        &   Yes        & (No)  &   \denning     & $21\pm 1.9$  &  $\lessapprox 158$ &  $\lessapprox 9\cdot 10^{-8}$\\	[0.5ex]
1979-06-26  &  S                  & $39^{+26}_{-15}$   &  $6.9^{+41}_{-5.9}\cdot 10^{-17}$  &  Possibly     &    No       & --  &  --           &      --      & -- &     \\	[0.5ex]
1980-01-02  &  E                  & $3^{+1}_{-2}$      &  $1.6^{+9.5}_{-1.4}\cdot 10^{-12}$ &  Possibly        &   Possibly         & (No)  &  \denning    & $21\pm 2.5$ & $\lessapprox 158$ & $\lessapprox 9\cdot 10^{-8}$ \\	
                                  &                    &                                    &               &              &     &               &              & & & \\[-2ex]
                                   \hline

           \end{tabular}
           
   \label{tab:particles}
\end{table*}

Similarly, for the case when two potential trail particles were  detected during the traverse of the trail of 
comet \hondaa 
in 1976, we get a  flux of
\begin{equation}
\Phi_{\mathrm{45P}} = \frac{N}{A} = \frac{2}{\mathrm{10\,days \cdot 0.003\,m^{2}}} = 67 \,\mathrm{m^{-2}\,day^{-1}}. \label{eq:phi_2}
\end{equation}
and a dust density of 
\begin{equation}
D_{\mathrm{45P}} =  2\cdot 10^{-8}\,\mathrm{m^{-3}} \label{eq:d_2}.
\end{equation}  
At first glance, these fluxes seem to be very high, and they are indeed three to four orders of 
magnitude larger than the values predicted by the model. 
One has to take into account, however, that the model predicts the fluxes of particles about
$\mathrm{100\, \mu m}$ in size and bigger while Helios detected particles at least a factor of ten smaller, 
and our flux estimates refer to these  
smaller particles. If we assume a dust size distribution
following a power law with a differential exponent of approximately $-4$ 
\citep[][references therein]{agarwal2010} and extend it to the approximately $\mathrm{10\, \mu m}$ 
 particles  as implied  by the Helios measurements, 
our derived flux values are in reasonable agreement with the simulated fluxes. 

We do not consider any statistical uncertainty here because systematic effects  most likely lead
to much larger uncertainties. For example, our calculation assumes that the particles were detected 
with a sensor orientation represented by the maximum of the sensor area $A$. Figure~\ref{fig:directions}, however,
shows that this is most likely not the case for most of the detections. Instead detections close to 
the edge of the field-of-view are more likely, which would imply significantly higher fluxes. 
On the other hand, dust sensitive sensor side walls would increase the sensitive area and reduce the
derived fluxes (Section~\ref{sec:discussion}). Other factors are the uncertainty in the spatial extent 
of the  trail 
and the identification of trail particles in the Helios data. In conclusion, we expect that the 
dust flux estimates we performed here have an uncertainty of at least a factor
of  ten. 

Dust fluxes simulated by IMEX can generally be considered as lower limits, for two reasons. First, 
the  model simulates only particles larger than $\mathrm{10^{-8}\,kg}$ (corresponding to
approximately $\mathrm{100\, \mu m}$), while the cometary 
trails most likely contain smaller particles as well \citep{agarwal2010}. Given that such
smaller particles are more susceptible to radiation pressure and Poynting-Robertson drag 
than the larger ones 
such particles get dispersed from the comet's orbit faster than the 
larger trail particles, however, a fraction of the recently released small particles remains close to 
the  trail for some time (cf. Section~\ref{sec:discussion}). Second, by comparing model results 
with cometary meteoroid stream observations, \citet{soja2015a} concluded that the model 
likely underestimates the true fluxes of $\mathrm{100 \mu m}$ and bigger particles 
by at least an order of magnitude. 

\section{Discussion}

\label{sec:discussion}

Our simulations give the best agreement with the particle detected by Helios in 1978. It may  be 
a  trail particle released from comet \denning. Three more particle detections show marginal
agreement with a cometary trail origin from comets \denningg (detection in 1980) and 
\hondaa (two detections in 1976). 
In Figure~\ref{fig:directions} the big crosses indicate the 
fields-of-view of the Helios dust instruments,  and these three impacts may have occurred close
to the edge of the sensor field-of-view. The crosses represent 
the sensor targets including shielding by the spacecraft structure 
\citep{gruen1980a}. The analysis of 
data obtained with the dust detectors on board the Galileo and Ulysses spacecraft, which were 
 impact ionization dust detectors of a design similar  to the Helios instruments
 except that they did not have the capability to measure time-of-flight spectra,
showed that their sensor side walls were sensitive to dust impacts as well \citep{altobelli2004a}. 
The sidewalls of the Helios sensors were made of metal, implying that they had a high yield for 
impact ionization too, although this was never tested in the laboratory. Dust-sensitive sidewalls would 
increase the sensor field-of-view and reduce the derived dust fluxes. The larger field-of-view would
give much better agreement of the 1976 and 1980 dust impacts with a cometary trail origin.

Figure~\ref{fig:fluxhigh} shows an offset in the dust particle detections as compared to the
times predicted by the model. This is particularly evident for the detection in 1978 which is offset 
from the time interval with the predicted highest dust fluxes by more than four days, corresponding
to a spacecraft motion of approximately 0.04~AU. Such an offset 
can have various reasons:

First, the IMEX model simulates only particles with masses $m=\mathrm{10^{-8}\,kg}$ and
bigger, corresponding to particle radii above approximately $\mathrm{100\, \mu m}$.  
The masses of the seven detected particles as derived from the calibration of the Helios dust instrument 
are 
at least {\em four orders of magnitude smaller}, 
corresponding to particle radii of a few micrometers to about $\mathrm{10\, \mu m}$ (cf. Table~\ref{tab:particles}). 
Such
smaller particles are more susceptible to radiation pressure and Poynting-Robertson drag 
than the larger ones. For particles larger than approximately $\mathrm{10\, \mu m}$ radius the 
ratio of the force of solar radiation pressure $F_{\rm rad}$ over that of gravity $F_{\rm grav}$ 
is below $\beta \lesssim 0.05$ for most materials, while micrometer-sized and smaller particles may have 
values of $\beta > 0.5$, and for some materials (e.g. metals) $\beta$ can be even larger than one \citep{burns1979a,kimura1999a}. 
It indicates that the measured particles were  more susceptible to radiation pressure and 
Poynting-Robertson drag than the ones simulated by the model. For example, the perihelion distance of a 
$\mathrm{10\,\mu m}$ particle (with $\beta=0.05$) on an eccentric orbit with semi-major axis $a=3\,\mathrm{AU}$ 
and eccentricity $e=0.7$ decreases by only approximately 0.0005~AU within 100~years, while for a  $\mathrm{100\,\mu m}$
particle this drift is ten times smaller. 
It indicates that Poynting-Robertson drag alone cannot account for a significant particle drift on the
time scales covered by the model. 
Second, the model uses a dust ejection model  to simulate
the dust emission from the comet nucleus due to water ice sublimation \citep[][see also \citet{soja2015a}]{crifo1997}.  
$100\,\mathrm{\mu m}$ particles (with density $\mathrm{1000\,kg\,m^{-3}}$) are ejected from 
the subsolar point on the surface of a nucleus with 1~km radius  at 0.7~AU 
heliocentric distance with about $80\,\mathrm{m\,s^{-1}}$. This is  well above the escape speed from the 
nucleus. Smaller particles have higher ejection speeds. Although the detailed particle motion is also 
strongly affected by the ejection direction from the nucleus surface as well as solar radiation pressure and Poynting-Robertson 
drag, the particle ejection due to sublimation adds to the particle drift in the vicinity of the nucleus. 
Third, the model simulates the particle dynamics for only up to 300 years, depending on the comet's
orbital period. This limitation was necessary based on the  computing power available at the time 
in order to simulate the dynamics of a sufficiently large number of dust particles for all 420 comets covered by the
model \citep{soja2015a}. The previous considerations show that  older particles are likely to be detected 
also further away from the nucleus. 
Finally, with increasing time  encounters with planets, in particular Jupiter, may significantly 
disturb the dust trails and move particles away from their source comet. One or more of these effects may 
explain the difference between the measured and the predicted particle detection time. 

Particle masses given in Table~\ref{tab:particles} were derived from the calibration of the 
Helios dust instruments in the laboratory. Because of the venetian blind-type 
target there is a large spread in the recorded impact charge depending on where the impact occurred, on the 
entrance side or on the multiplier side of the target strip. Furthermore, impacts on the sensor side-wall may 
generate a very reduced impact charge. Therefore, the size (mass) of the impactors may be significantly 
underestimated, even more so if cometary particles are fluffy aggregates as indicated, for example, by the
results from the Rosetta mission to comet 67P/Churymov-Gerasimenko \citep{guettler2019,kimura2020a,kimura2020b}. 
This is also supported by the particle dynamics: Particles released at perihelion from comets \denningg and \honda, 
whose orbital eccentricity is $e = 0.82$, cannot remain in a bound heliocentric orbit unless their $\beta$
ratios are smaller than 0.09. This means that particles approximately $\mathrm{5\mu m}$ in radius and 
smaller are very quickly removed from these trails and escape from the solar system on hyperbolic trajectories. 
For very porous particles with 85\% porosity, consistent with Rosetta results,
this limit increases to $\mathrm{3.0 \cdot 10^{-11}\,kg}$, corresponding to a particle radius 
of about $\mathrm{14\mu m}$.
Therefore, if Helios really detected 
particles belonging to a cometary trail, they must have been significantly {\em bigger} than the  
sizes derived from the instrument calibration. Thus, the chances are low that Helios observed rather micrometer 
sized cometary trail particles. This is similar to the situation when Ulysses detected Jupiter stream 
particles \citep{gruen1998}: We only learned from modelling that the particles  actually had nanometer 
sizes, i.e. they were much smaller than particle sizes derived from the instrument calibration.

Remarkably, between 1881 and 2014, comet \denningg was observed only during its 1978 apparition.
During all the other apparitions it was not re-discovered, although, based on the observing conditions 
and predicted brightnesses, recoveries 
should have been possible a few times 
\citep{beech2001,sato2014}. From these non-detections, the authors concluded that this comet may not 
have been active during 
all of its past apparitions. 
If \denningg  indeed had dormant periods in the past, the dust spatial
density or the volume filled by the trail, or both, may be overestimated by the IMEX model.
In a similar way, comet \finlayy may also be
evolving into a transitional asteroid-like object \citep{beech1999}, although it possess 
the ability for repetitive energetic outbursts \citep{ishiguro2016}.

The detections of our seven candidate cometary trail particles  close to Venus's orbit is
intriguing (cf. Figure~\ref{fig:orbitplot}). In addition to the in-situ dust instruments,   
Helios also carried a zodiacal light photometer 
which discovered a heliocentric dust ring along Venus' orbit \citep{leinert2007}. This ring
was later confirmed by observations with the Heliospheric Imager instruments 
on board the two STEREO spacecraft \citep{jones2013b}, and  in-situ  measurements by the
Arrayed Large-Area Dust Detectors in INterplanetary space (ALADDIN) on board the IKAROS
spacecraft show a dust flux variation that may be connected with a Venusian dust ring \citep{hirai2014,yano2014}. 
From the STEREO observations, the enhancement 
in the dust spatial density in the Venus  ring with respect to the
interplanetary dust background   was found to be only 8\% at most \citep{jones2017}. Interesting 
enough, the STEREO observations showed a step-like increase in the dust density on the inner side 
of Venus' orbit while there was no drop in dust density detected on the outer side. 
Furthermore, dynamical modelling indicates that relatively small particles as measured by Helios 
($\mathrm{\lesssim 10\, \mu m}$) cannot be effectively trapped in resonances with Venus due to the stronger Poynting-Robertson
drag and thus are unlikely to contribute to a dust enhancement in the Venus ring \citep{pokorny2019,sommer2020}. 
The relatively weak enhancement in dust density together with the required large particle sizes 
makes it unlikely that at $\eta = 135^{\circ}$ Helios detected impacts by Venus dust ring particles, 
although their impact speed and directions are in the same range as  those of the cometary trail
particles (Figure~\ref{fig:directions}; $v_{\mathrm{imp}} = \mathrm{17\,km\,s^{-1}}$; 
$\lambda_{\mathrm{ecl}} = 8^{\circ}$, $\beta_{\mathrm{ecl}} = -5^{\circ}$) 
 
Figure~\ref{fig:orbitplot} reveals the remarkable coincidence that at a true anomaly angle of $\eta \approx 135^{\circ}$
Helios traversed three known cometary trails. Only because of this coincidence and high dust fluxes in the trails,
\citet{altobelli2006} were able to identify a concentration of seven dust impacts in the data of 
the relatively small Helios detectors. Figure~\ref{fig:flux} shows that if the spacecraft had 
traversed  only one of the trails of \hondaa or \denningg but not both at $\eta \approx 135^{\circ}$, 
Helios would only have reported three or four particle detections, respectively, at this 
true anomaly angle. It is unlikely that \citet{altobelli2006} would have recognised such a smaller number of particles 
as potentially being of cometary trail origin. 

By an interesting coincidence, from studying the orbital data of 
fireballs associated with the $\alpha$ Capricornids 
meteor stream in Earth's atmosphere \citet{hasegawa2001} concluded that exactly our three candidate 
comets \honda, 
\denningg and \machholz, among three additional comet and asteroid candidates, could be associated with this meteor stream. 
Comet \hondaa   
 was also predicted to be the source of a meteor shower in the Venus atmosphere  \citep{vaubaillon2006}.


The clustering of trail crossings at a true
anomaly angle of $\eta \approx 135^{\circ}$ also explains why there was no particle 
concentration detected at  approximately $\eta \approx 225^{\circ}$ even though the model
predicts trail crossings here as well (cf. Figure~\ref{fig:flux}): First, the fluxes predicted 
by IMEX for single trails are somewhat lower than
for the traverses at $\eta \approx 135^{\circ}$ and second, the cometary orbits are much 
more widely dispersed in space (Figure~\ref{fig:orbitplot}).

Comet trails were first identified in the IRAS all sky survey \citep{sykes1992}. Subsequently, 
in a survey of 34 Jupiter-family comets with the Spitzer Space Telescope,  at least 80\% of the comets 
 were associated with dust trails \citep{reach2007}. Comet trails were also studied with 
 the Diffuse Infrared Background Experiment (DIRBE)
 instrument on board the Cosmic Background Explorer 
 \citep[COBE;][]{arendt2014} and with ground-based observations 
 in the visible range \citep{ishiguro2007}. 
 Unfortunately, none of the comets identified in our analysis 
 was contained in any of these surveys.
 
Radar observations of cometary comae can provide information about the particle sizes comprising
 the coma. Observations with the Arecibo Observatory planetary radar system showed that 
the coma of  comet \hondaa  contains particles larger than 2~cm  \citep{springmann2017},
while the existence of smaller particles could not be excluded. 

Particle mass spectra can provide valuable information about the composition and evolution of the 
source bodies of the detected particles. Only 
two interplanetary dust particles were successfully analysed with the  Cosmic Dust Analyzer \citep[CDA;][]{srama2004} 
on board the Cassini spacecraft during
 its journey to Saturn \citep{hillier2007}. Surprisingly, both particles had a very similar metallic (iron)
composition with an absence of typical features expected for silicate minerals (e.g. silicon). The authors 
concluded that the particles were compatible with an asteroidal origin, although an origin  
from Jupiter family comets would also be possible. A few of the Helios spectra shown in Figure~\ref{fig:spectra} also
show a broad feature covering iron (56~amu). In particular, the mass spectrum of the particle detected in 1979 
is very similar to the CDA detections. Furthermore, at least five additional Helios particles (both in 1976, 1977, 1978 and 1980) 
show a broad peak covering silicon (28~amu),  compatible with the presence of silicates. 

\section{Future Perspectives}

\label{sec:perspectives}

Our analysis  shows that the identification of cometary trails in in-situ 
dust data may be possible even with a relatively small dust instrument. It opens the perspective 
to identify impacts of cometary trail particles in the data sets of other space missions as well.
The Ulysses spacecraft provided the longest continuous data set of in-situ dust measurements in interplanetary
space presently available: Dust measurements were
collected during 17 years while the spacecraft made three revolutions around the Sun   \citep{gruen1992b,krueger2010b}. 
We may be able to 
identify impacts of cometary trail particles in this data set as well because the spacecraft  traversed 
the same regions of space up to three times. Given that the sensitive area of the Ulysses dust detector was 
about a factor of eight larger than the combined area of the 
Helios detectors, the search for cometary meteoroid trails in the Ulysses data set is promising and ongoing. 
Finally, the dust detector on board the New Horizons spacecraft \citep{horanyi2008,poppe2010} may reveal 
cometary trail crossings in the outer solar system. 

Large variations in the predicted dust fluxes from comet to comet have to be expected in the IMEX model
because the ejection velocity, 
mass distribution, and dust production rate -- all parameters of the model potentially 
as a function of time -- likely vary for each comet and are not well constrained for 
many comets yet. This may be improved in the future for the comets found in our 
analysis to yield more reliable flux predictions. 

The present IMEX model has a lower particle mass limit of $\mathrm{10^{-8}\,kg}$.
Future model 
extensions may include smaller particle sizes as well to cover
 particles that are more susceptible to solar radiation pressure. The trajectories of such smaller 
particles are expected to be offset from those of their
bigger counterparts. Future
simulations with such an extended model may give better agreement for the comets identified in our present analysis, 
and they may  reveal additional comets to explain the Helios trail particle detections.

Many meteor streams and  fireballs observed in the Earth's atmosphere were successfully 
traced back to their parent comets and asteroids \citep[e.g.][]{jenniskens2006}. With high flying aircraft 
extraterrestrial dust was collected in the Earth atmosphere and its analysis in the laboratory 
provided a wealth of information \citep[][and references therein]{jessberger2001}, however, their 
individual source bodies usually remain
unknown. Only in very rare cases could "targeted" collections catch particles from a dedicated comet 
when the Earth crossed its trail, e.g.  comet 26P/Grigg-Skjellerup \citep{busemann2008,davidson2012}. There have also 
been attempts to measure the particle composition 
of the induced meteors in the Earth atmosphere by ground-based observations 
\citep[e.g.][]{toscano2013}, however, these are strongly limited by contamination from 
atmospheric constituents. The in-situ detection and analysis of cometary trail particles in space  opens a new 
window to remotely measure the composition of  celestial bodies without the necessity to fly a spacecraft to the source 
objects.

The DESTINY$^+$ (Demonstration and Experiment of Space Technology for INterplanetary voYage with Phaethon fLyby and dUst Science) 
mission will be launched to the active near-Earth asteroid
(3200) Phaethon in 2024 \citep{kawakatsu2013,arai2018}. 
The DESTINY$^+$ Dust Analyzer \citep[DDA;][]{kobayashi2018b} on board is an upgrade of CDA  which very successfully 
investigated dust throughout the Saturnian system \citep{srama2011}. The instrument will measure the 
composition of interplanetary and interstellar dust 
during the spacecraft's interplanetary journey to Phaethon as well as dust released from Phaethon during a
close flyby at the asteroid. Recently, Phaethon's dust trail was identified in optical images obtained with the
STEREO spacecraft \citep{battams2020}. 

We also performed IMEX simulations for the DESTINY$^+$  mission 
 based on the spacecraft trajectory presently available. Our results show that DESTINY$^+$ will 
traverse the trail of comet \hondaa  three times in 2026 and 2027. If this is confirmed in the future for 
the real  DESTINY$^+$ trajectory to be 
flown in space, 
this coincidence provides the interesting opportunity to compare the 
Helios spectra of likely trail particles  with the  high-resolution DDA spectra from the same source comet. 
Such a comparative study may also give new insights into the interpretation of the full set of 235 Helios 
mass spectra.

Other present or future space missions equipped with dust detectors include BepiColombo which has 
 the Mercury Dust Monitor on board \citep[MDM,][]{kobayashi2020}. Even though MDM is partially obstructed by the 
heat shield of BepiColombo during the spacecraft's interplanetary passage to Mercury, the sensor may be able to detect 
particles in the trail of comet
2P/Encke en-route to  planet Mercury. The Martian Moons Exploration (MMX) mission will be launched to
Phobos and Deimos in 2024, and the large area ($\mathrm{\sim 1\,m^2}$) dust impact detector 
on board  may detect cometary trails en-route to Mars \citep{kobayashi2018a,krueger2019c}. Furthermore, 
we encourage the Europa mission team with its dust telescope SUDA \citep{kempf2018} to search for interplanetary 
dust and possible trails along its way to Jupiter. Finally, the JUICE mission \citep{witasse2019} will not carry 
a dust sensor, but the radio plasma instrument might allow the detection of interplanetary dust particles. 
Plasma wave instrument activation during dust trail crossing zones is recommended.

\section{Conclusions}

\label{sec:conclusions}

We have re-analysed a subset of seven dust impacts measured in the 1970s by the Helios dust instruments in 
the inner solar
system.  The particles were originally identified by \citet{altobelli2006} as potential cometary trail particles 
due to their clustering in a small region of space at a true anomaly angle of $135\pm 1^{\circ}$ during several 
revolutions of Helios around the Sun. We 
have modelled Helios traverses of cometary meteoroid trails with the 
Interplanetary Meteoroid Environment for eXploration (IMEX) dust streams in 
space model \citep[][]{soja2015a}. The model simulates recently created cometary meteoroid streams 
in the inner solar system. 

The identification of potential cometary trail particles in the Helios data greatly benefitted from the 
clustering of trail traverses in a rather narrow region of space. 
We identified comets 
45P/Honda-Mrkos-Pajdu\v{s}{\'a}kov{\'a} and 72P/Denning-Fujikawa as the likely sources for a subset of 
four of the candidate trail particles. By using the Helios measurements in combination with the simulation results 
we found  spatial densities of approximately $\mathrm{10\, \mu m}$  dust particles in the trails of these comets
 of  $\mathrm{10^{-8}\,m^{-3}}$ to $\mathrm{10^{-7}\,m^{-3}}$. 
Our analysis shows that trail particles 
are likely detectable with an in-situ dust impact detector when the spacecraft 
traverses such a dense cometary dust trail. It opens a new 
window to remote compositional analysis of  celestial bodies without the necessity to fly a spacecraft close 
to or even land on the source objects.

\section*{Acknowledgements}
 
The IMEX Dust Streams in Space model was developed unter ESA funding (contract 4000106316/12/NL/AF - IMEX). 
We are grateful to the 
MPI f\"ur Sonnensystemforschung, Chiba Institute of Technology and the University of Stuttgart for their support. 
 H. Kimura gratefully acknowledges support by the Grants-in-Aid for Scientific Research (KAKENHI number 19H05085) of the 
Japan Society for the Promotion of Science (JSPS).  
 We thank an anonymous referee whose comments  improved the presentation of our results. 


\end{document}